\documentclass[aps,prx,twocolumn,10pt,amsmath,amssymb,floatfix,superscriptaddress]{revtex4-2}

\usepackage{graphicx} 
\usepackage{bm} 
\usepackage{hyperref}
\usepackage{braket,bbold,color}
\usepackage{hyperref}
\usepackage{multirow}
\usepackage[table]{xcolor}
\newcommand{\bs}[1]{\boldsymbol{#1}}

\begin{document}

\title{Berry Curvature of Low-Energy Excitons in Rhombohedral Graphene}

\author{Henry Davenport}
\affiliation{Blackett Laboratory, Imperial College London, London SW7 2AZ, United Kingdom}

\author{Frank Schindler}
\affiliation{Blackett Laboratory, Imperial College London, London SW7 2AZ, United Kingdom}

\author{Johannes Knolle}
\affiliation{Technical University of Munich, TUM School of Natural Sciences, Physics Department, 85748 Garching, Germany}
\affiliation{Munich Center for Quantum Science and Technology (MCQST), Schellingstr. 4, 80799 M\"unchen, Germany}
\affiliation{Blackett Laboratory, Imperial College London, London SW7 2AZ, United Kingdom}

\begin{abstract}
We investigate low energy excitons in 
rhombohedral pentalayer graphene encapsulated by hexagonal boron nitride (hBN/R5G/hBN), focusing on the regime at the experimental twist angle $\theta = 0.77^\circ$ and with an applied electric field. We introduce a new low-energy two-band model of rhombohedral graphene that captures the band structure more accurately than previous models while keeping the number of parameters low. Using this model, we show that the centres of the \emph{exciton} Wannier functions are displaced from the moiré unit cell origin \emph{by a quantised amount} -- they are instead localised at $C_3$-symmetric points on the boundary. 
We also find that the exciton shift is electrically tunable: by varying the electric field strength, the exciton Wannier centre can be exchanged between inequivalent corners of the moiré unit cell. 
Our results suggest the possibility of detecting excitonic corner or edge modes, as well as novel excitonic crystal defect responses in hBN/R5G/hBN. Lastly, we find that the excitons in hBN/R5G/hBN inherit excitonic Berry curvature from the underlying electronic bands, enriching their semiclassical transport properties. Our results position rhombohedral graphene as a compelling tunable platform for probing exciton topology in moiré materials.

\end{abstract}

\maketitle
\section{Introduction}
The study of Berry curvature effects has played a central role in advancing our understanding of non-interacting electrons in condensed matter systems~\cite{BerryPhaseEffects}. More recently, there has been growing interest in extending this framework to interacting systems. Efforts in this direction include the topological classification of ground states in the presence of interactions~\cite{IntGroundstate1, IntGroundstate2, IntGroundstate3, IntGroundstate4}, as well as the investigation of Berry curvature effects and the topological classification of interaction-induced excitations, such as excitons~\cite{ExcitonTopologyDavenport, ExcitonTopologyPaperKwan, ExcitonTopologyPaperTitus, ExcitonTopologyPaperQian, ExcitontopologySlager, ExcitonTopologyPaperUchoa}.

\begin{figure}[ht]
    \centering
\includegraphics[width=1.0\linewidth]{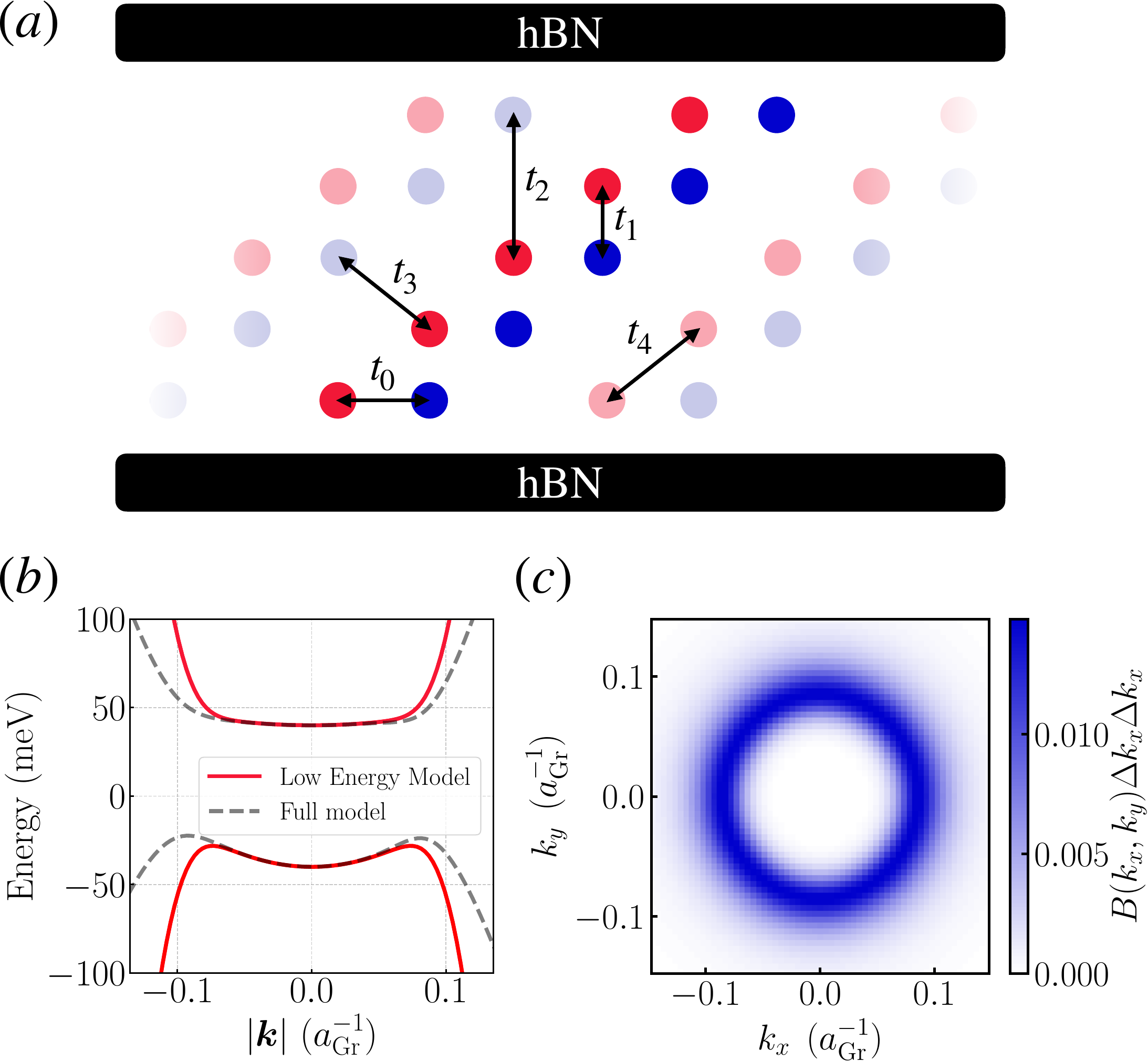}
    \caption{Rhombohedral graphene encapsulated between two hBN layers. The hopping parameters of our tight-binding model are labelled $t_{0\dots 4}$ in $(a)$. The $A$ sublattice of graphene is shown in red and the $B$ sublattice in blue.
    $(b)$ Dispersions of the low energy model and the full model of the free-standing rhombohedral graphene (with displacement field $u_D = 20\mathrm{meV}$). $(c)$ The Berry curvature [$B(k_x, k_y)$] distribution of the low energy model.}
    \label{fig:combinedFigure}
\end{figure}

Rhombohedral graphene has emerged as a compelling platform for studying non-trivial ground-state topology induced by interactions. Experiments have found that rhombohedral graphene exhibits both the integer and fractional quantum Hall effects~\cite{RhombohedralGrapheneExperiment} and a comprehensive theoretical understanding of these phenomena remains an area of active research~\cite{BernevigRhombohedral, VishwanathPaper, AshvinPaper2, bernevig2025berrytrashcanmodelinteracting, yu2024moirefractionalcherninsulators, SenthilPaper, QuantumAnomalousHallSenthil, QuantumAnomalousHallEffectZhang}. In this work, we take an alternative approach and focus on the \emph{excitations} rather than the ground state, specifically the excitons at charge neutrality. We study rhombohedral pentalayer graphene (R5G) encapsulated by hexagonal boron nitride (hBN), shown in Fig.~\ref{fig:combinedFigure}a. When an electric field is applied we can obtain two well-gapped electronic bands around the Fermi level at charge neutrality~\cite{BernevigRhombohedral}. Since the bands immediately above and below the Fermi energy are layer polarised, the twist angle of the hBN either side of the R5G strongly influences the topology and dispersion of these electronic bands. The electronic bands are therefore highly tunable making hBN/R5G/hBN an ideal testbed for studying the topology of low energy excitons emerging from topological electron bands.

In this paper we show that, at the experimentally relevant twist angle $\theta = 0.77 ^\circ$~\cite{RhombohedralGrapheneExperiment}, the exciton Wannier states are shifted to the edge of the moiré unit cell and their Wannier centre depends on the electronic potential applied across the rhombohedral graphene layers. They are therefore analogous to obstructed atomic insulators in non-interacting electron systems~\cite{TQCPaper}. We note however that this shift is inherited from the electronic bands and so is not like the shift excitons described in Ref.~\onlinecite{ExcitonTopologyDavenport} where the excitons have maximally localised exciton Wannier centres~\cite{jonahMLXWF, davenport2025excitonberryology} shifted with respect to the electronic Wannier centres. Despite this, we expect that rhombohedral graphene still exhibits novel `excitonic' defect and edge responses. We also find that the Berry curvature of the excitons can be tuned using the applied electric field, allowing for an experimental exploration of the role of Berry curvature on excitonic properties. Given these attributes, rhombohedral graphene presents a unique system to explore the effects of exciton Berry curvature on exciton transport properties, both experimentally and theoretically.

This paper is structured as follows. We begin by outlining the non-interacting physics underlying the band dispersion and Berry curvature distribution. In Sec.~\ref{sec:NonInteractingBandStructure} we introduce a new simple low-energy 2 band model of rhombohedral graphene that better captures the dispersion shape around the Dirac point compared to the well-studied $k^5$ model~\cite{SenthilPaper, lowEnergyModelWannier}. In addition, we show how the moiré potential modifies the dispersion and band topology. In Sec.~\ref{Sec:Excitons} we detail our methodology for calculating the exciton dispersion and exciton Berry curvature. Our findings indicate that the applied electric field can be used to tune the lowest-energy excitons to regimes where the exciton Berry curvature is enhanced. We predict that this could lead to a measurable thermal Hall effect~\cite{zhang2024thermal}. Additionally, we identify regimes where the exciton Wannier centres are obstructed from the centre of the moiré unit cell, with the Wyckoff position they are centred at tunable using the applied electric field. In Sec.~\ref{sec:Discussion} we discuss the experimental consequences of this exciton shift in terms of defect responses detectable using Electron Energy Loss Spectroscopy (EELS)~\cite{ExcitonImage1, ExcitonImage2}. 

\section{Band structure}
\label{sec:NonInteractingBandStructure}
We first derive an economical model that captures the band structure of R5G as well as that of hBN/R5G/hBN with a moiré twist.

\subsection{Low energy model}
We begin by formulating a new low-energy effective model for isolated pentalayer rhombohedral graphene, which serves as the basis for all subsequent analysis. Our starting point is the full 10-band tight-binding model previously studied in Ref.~\onlinecite{VishwanathPaper}, with interlayer and intralayer hopping parameters as illustrated in Fig.~\ref{fig:combinedFigure} (see Appendix~\ref{apdx:FullModelAndLowEnergyModel} for a full description). This model includes an external perpendicular electric field - known as the displacement field. To model this we add a layer-dependent onsite potential, where the potential difference between adjacent layers is uniform and denoted by $u_D$~\cite{VishwanathPaper}. 

When $u_D$ is non-zero, and for momenta near the Dirac point, the electronic wave functions associated with the conduction and valence bands closest to the Fermi level become strongly localised on sublattice $A$ of layer 1 and sublattice $B$ of layer 5~\cite{VishwanathPaper}. This layer polarization enables a projection onto the corresponding subspace, yielding a reduced two-band model that captures the low-energy physics at charge neutrality (see Appendix~\ref{apdx:FullModelAndLowEnergyModel} for detailed derivations). With our method we can systematically include higher order corrections beyond the simpler two band models studied previously~\cite{SenthilPaper}. In particular we incorporate a key correction absent from previous work: namely, the intrinsic quadratic dispersion that dominates at small momenta $\boldsymbol{k}$ near the Dirac point [see Fig.~\ref{fig:combinedFigure} panel (b)]. We show that this curvature can be systematically retained within the effective theory by including a correction quadratic in the momentum. This results in the following two-band Hamiltonian that more accurately reflects the low-energy behaviour of rhombohedral graphene,
\begin{equation}
\begin{aligned}
    &h_{\mathrm{eff}}(\boldsymbol{k}) = \\ &\left(
\begin{array}{cc}
 2 u_D+\gamma_-(k_x^2+k_y^2) & \frac{v_F^5}{t_1^4}\left(k_x-i k_y\right)^5 \\
 \frac{v_F^5}{t_1^4}\left(k_x+i k_y\right)^5 & -2 u_D+\gamma_+(k_x^2+k_y^2) \\
\end{array}
\right),
\label{eq:2by2Model}
\end{aligned}
\end{equation}
where $v_F = \frac{\sqrt{3}}{2}t_0 a$ and $a$ is the graphene lattice constant. The numerical values of the hoppings are given in Appendix~\ref{apdx:FullModelAndLowEnergyModel}~\cite{VishwanathPaper, MoirePotential}. The $\gamma_{\pm}$ take a simple form if we assume small $u_D$ (specifically $u_D \ll t_3 \ll t_0$) and expand to first order in $u_D$,
\begin{align}
\gamma_{\pm} &\approx \frac{2 v_F  v_4}{t_1}\pm\frac{u_D v_F^2}{t_1^2},
\label{eq:gammaCorrections}
\end{align}
where $v_4 = \frac{\sqrt{3}}{2}t_4 a$. This simplified model predicts that, for $u_D > 0$, the curvature of the higher-energy band's dispersion (at the Dirac point) changes sign when the displacement field reaches $u_D = \frac{2v_4 t_1}{v_F} \sim 35 \:\mathrm{meV}$. In contrast, the curvature of the lower-energy band at the Dirac point remains unchanged for all \( u_D \geq 0 \). This sign change is absent in the model presented in Ref.~\cite{SenthilPaper}, which neglects this diagonal quadratic correction.

Fig.~\ref{fig:combinedFigure}b compares the band dispersions of the full and simplified models for a representative set of parameters. As expected, both models closely agree near the Dirac point, with increasing deviations at larger momenta. In addition to capturing the band dispersion, the simplified model also reproduces the Berry curvature distribution of the full model~\cite{SenthilPaper}. As shown in Fig.~\ref{fig:combinedFigure}c, the Berry curvature in the simplified model is concentrated in a ring around the Dirac point. This ring occurs at the momenta where the band dispersion transitions from being nearly flat to rapidly increasing or decreasing.

\subsection{Moiré Hamiltonian}

\begin{figure}
    \centering
    \includegraphics[width=1.0\linewidth]{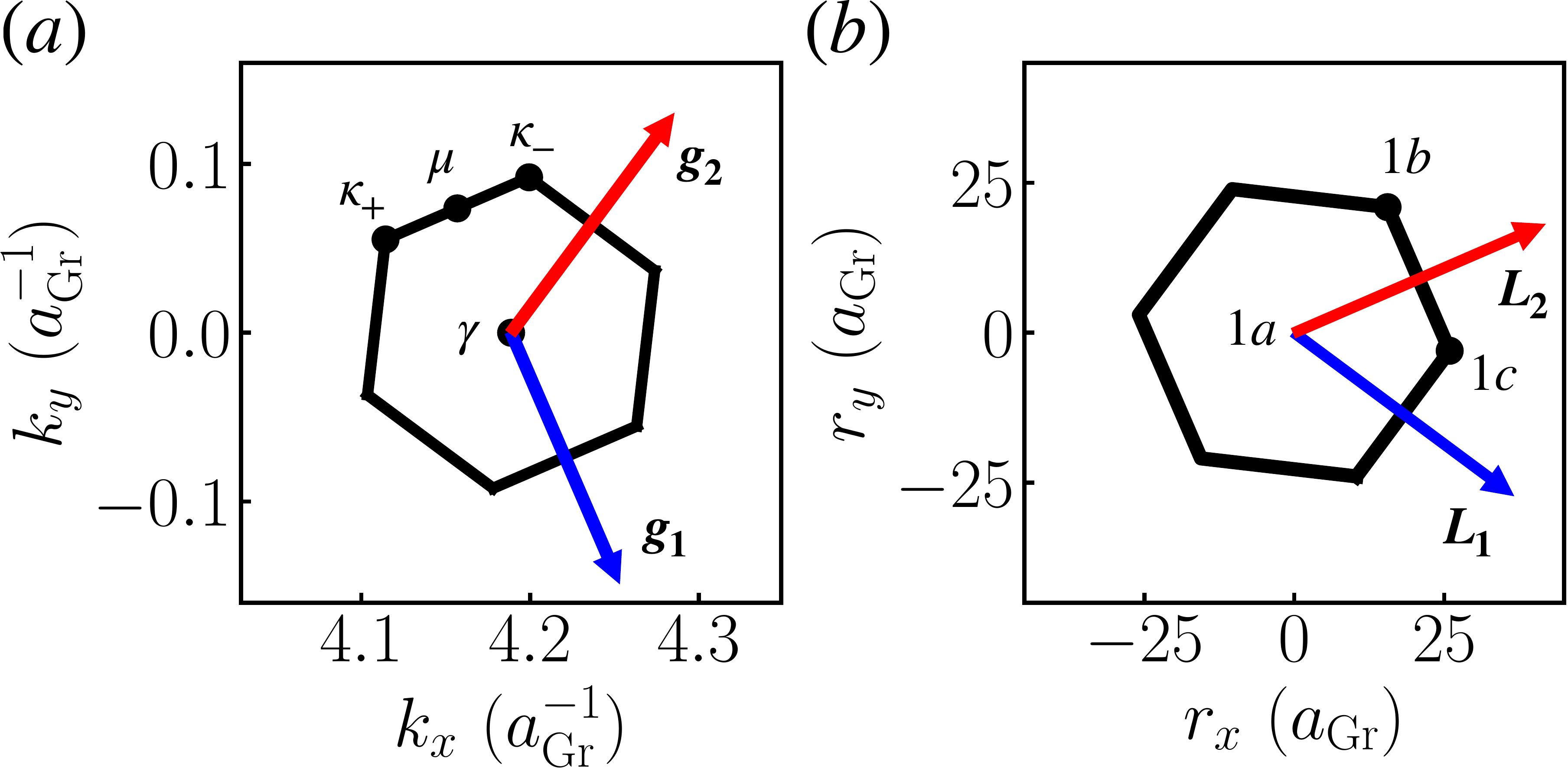}
    \caption{The mini Brillouin zone $(a)$ and unit cell $(b)$ for hBN/R5G/hBN with twist angle $\theta = 0.77^\circ$. The moiré reciprocal lattice vectors are labelled $\boldsymbol{g_1}, \:\boldsymbol{g_2}$, and the high symmetry points are $\gamma$, $\kappa_+$, $\kappa_-$ and $\mu$. The real space unit cell [panel $(b)$] shows the lattice vectors $\boldsymbol{L_1}, \: \boldsymbol{L_2}$ and maximal Wyckoff positions $1a,\: 1b, \: 1c$.}
    \label{fig:mBZandUnitCell}
\end{figure}

We investigate pentalayer rhombohedral graphene (R5G) encapsulated between two layers of hexagonal boron nitride (hBN), as illustrated in Fig.~\ref{fig:combinedFigure}. In contrast to most previous work, which has considered coupling of the R5G to hBN on only one side~\cite{VishwanathPaper, BernevigRhombohedral, yu2024moirefractionalcherninsulators}, we assume that both the top and bottom hBN layers couple equally to the R5G (as also studied in Ref.~\cite{BernevigRhombohedral}). Due to the lattice parameter mismatch between hBN and graphene, a moiré pattern forms which folds the bands into the mini BZ (mBZ) (shown in Fig.~\ref{fig:mBZandUnitCell}a). The (real space) unit cell is enlarged compared to the isolated R5G unit cell (Fig.~\ref{fig:mBZandUnitCell}b). The moiré unit cell has a periodicity ($a_{\mathrm{moir\acute{e}}} = |\boldsymbol{L_{1/2}}|$ in Fig.~\ref{fig:mBZandUnitCell}b) which depends on the relative twist angle $\theta$ between the primitive lattice vectors of the hBN and R5G and the lattice constant missmatch between graphene and hBN~\cite{MoirePotential}. We study the regime where both the top and bottom hBN layers are aligned at the same angle relative to the graphene, thereby avoiding the formation of a \emph{super-moiré} pattern. This double alignment is possible experimentally, for example in Ref.~\onlinecite{doubleAlignmentExample}, graphene is encapsulated in hBN with the two hBN sheets aligned. This is achieved by aligning the exfoliated straight edges of each flake using images from an optical microscope.
We model the hBN layers as an effective potential acting on the outermost layers of the rhombohedral graphene~\cite{MoirePotential}. Here we summarise the method used to model the moiré system, see Appendix~\ref{apx:MoireHamiltonian} for the full details. Since our effective Hamiltonian retains only the $A$ sublattice in the bottom layer and the $B$ sublattice in the top layer, we apply the moiré potential exclusively to these sites. For layer $l$ we use,~\cite{MoirePotential} 
\begin{equation} V^{l}_{\mathrm{eff}}(\boldsymbol{r}) = V_0 +  V_1 \sum_{\substack{\eta \in \{1, -1\} \\ j\in \{1, 2, 3\}}}
\exp\left[\mathrm{i}\eta\left(\boldsymbol{g}_j \cdot \boldsymbol{r} + \psi_{\xi}+\phi_l \right)\right]. 
\end{equation}
Here, the sum runs over moiré reciprocal lattice vectors~$\boldsymbol{g}_j$, using the ``first harmonic” approximation where we retain only the first shell of reciprocal vectors surrounding the first moiré Brillouin zone~\cite{MoirePotential, BernevigRhombohedral}. There is a phase shift $\psi_{\xi}$ which depends on the stacking configuration, for example, the $A$ sublattice of graphene may lie above either a boron atom ($\xi = 0$) or a nitrogen atom ($\xi = 1$), and this leads to differing phases for the effective potential~\cite{MoirePotential}. We restrict our study to $\xi = 0$ stacking for both the top and bottom layers (for which $\psi_{\xi} = 223.5^\circ$)~\cite{MoirePotential}. The layer dependent phase factor is only nonzero on the top layer, $\phi_l = \frac{2\pi N_l}{3}\delta_{l, 4}$~\cite{BernevigRhombohedral}. This phase accounts for the relative shift in the moiré potential between the $A$ sublattice on the bottom later and the $B$ sublattice on the top layer~\cite{BernevigRhombohedral}.

We define the fermionic creation operators, $c^\dagger_{\boldsymbol{k}, \boldsymbol{Q}, l} = c^\dagger_{\boldsymbol{k}+ \boldsymbol{Q}, l}$ where the momentum $\boldsymbol{k}$ is in the mBZ, $\boldsymbol{Q}$ are the moiré reciprocal lattice vectors, and $l$ is the layer index. The total moiré Hamiltonian is,
\begin{equation}
\begin{aligned}
H_M = &H_0 + H_V = \\ &\sum_{\boldsymbol{k} \in \mathrm{mBZ}} \sum_{\boldsymbol{Q}, \boldsymbol{Q}'}\sum_{l, l'}h^{(m)}_{(\boldsymbol{Q},l), (\boldsymbol{Q}', l')} (\boldsymbol{k}) c^\dagger_{\boldsymbol{k}, \boldsymbol{Q}, l} c_{\boldsymbol{k}, \boldsymbol{Q}', l'},
\end{aligned}
\end{equation}
where,
\begin{equation}
\begin{aligned}
&h^{(m)}_{(\boldsymbol{Q}, l), (\boldsymbol{Q}', l')}(\boldsymbol{k}) \\&= \left\{[h_{\mathrm{eff}}(\boldsymbol{k}+\boldsymbol{Q})]_{l, l'} + V_0 \delta_{l, l'}\right\}\delta_{\boldsymbol{Q}, \boldsymbol{Q}'} + V_{1}\sum_{\substack{\eta \in \{1, -1\} \\ j\in \{1, 2, 3\}}} \delta_{\boldsymbol{Q}, \boldsymbol{Q}' - \eta \boldsymbol{g}_j}.
\label{eq:MoireHamiltonian}
\end{aligned}
\end{equation}
Here the first term $h_{\mathrm{eff}}(\boldsymbol{k})$ represents the original \emph{low energy} rhombohedral graphene model without the moiré coupling. Although the low energy two-band model has a continuous rotational symmetry, the hBN moiré potential breaks this but retains $C_3$ symmetry. As shown in  Fig.~\ref{fig:mBZandUnitCell}, in the presence of the $C_3$ symmetry there are three high symmetry points: $\gamma$, $\kappa_+$ and $\kappa_-$. To obtain a finite matrix size, we truncate the set of $\boldsymbol{Q}$ vectors in the moiré Hamiltonian to those within a ring around the $\gamma$ point with~$|\boldsymbol{Q}| \le 5 [2\pi \cdot (a_{\mathrm{moir\acute{e}}})^{-1}]$.
\subsection{Non-interacting band structure}

\begin{figure}
\centering\includegraphics[width=1.0\linewidth]{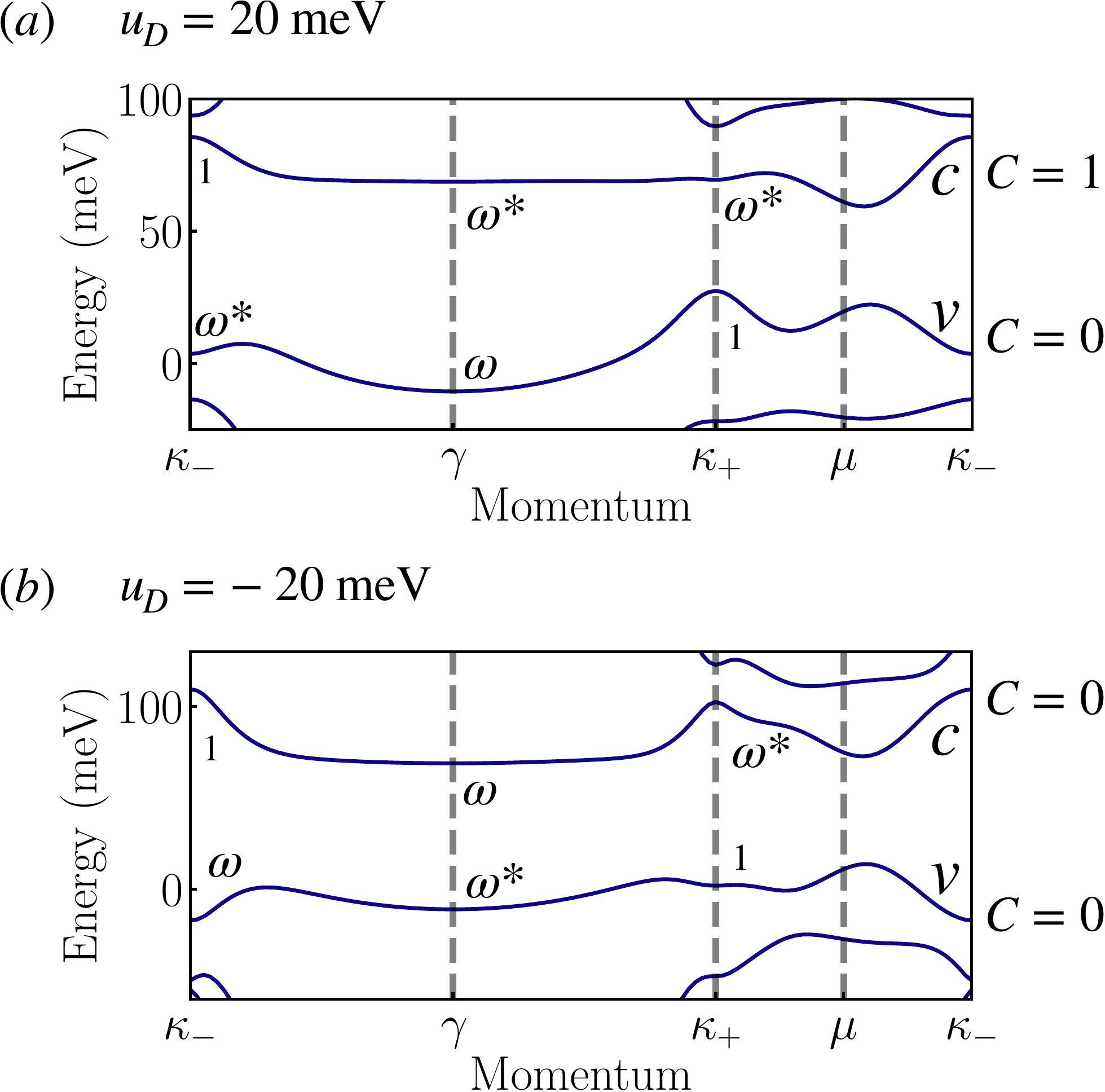}
    \caption{Electronic (non-interacting) dispersion at $\theta = 0.77^\circ$ for displacement field $u_D = 20 \:\mathrm{meV}$ and $u_D = -20 \:\mathrm{meV}$. The $C_3$ symmetry eigenvalues are shown at the high symmetry points in terms of $\omega = \exp \left(\mathrm{i} 2\pi/{3}\right)$.}
    \label{fig:nonIntDispersion}
\end{figure}

\begin{figure}
    \centering
    \includegraphics[width=1.0\linewidth]{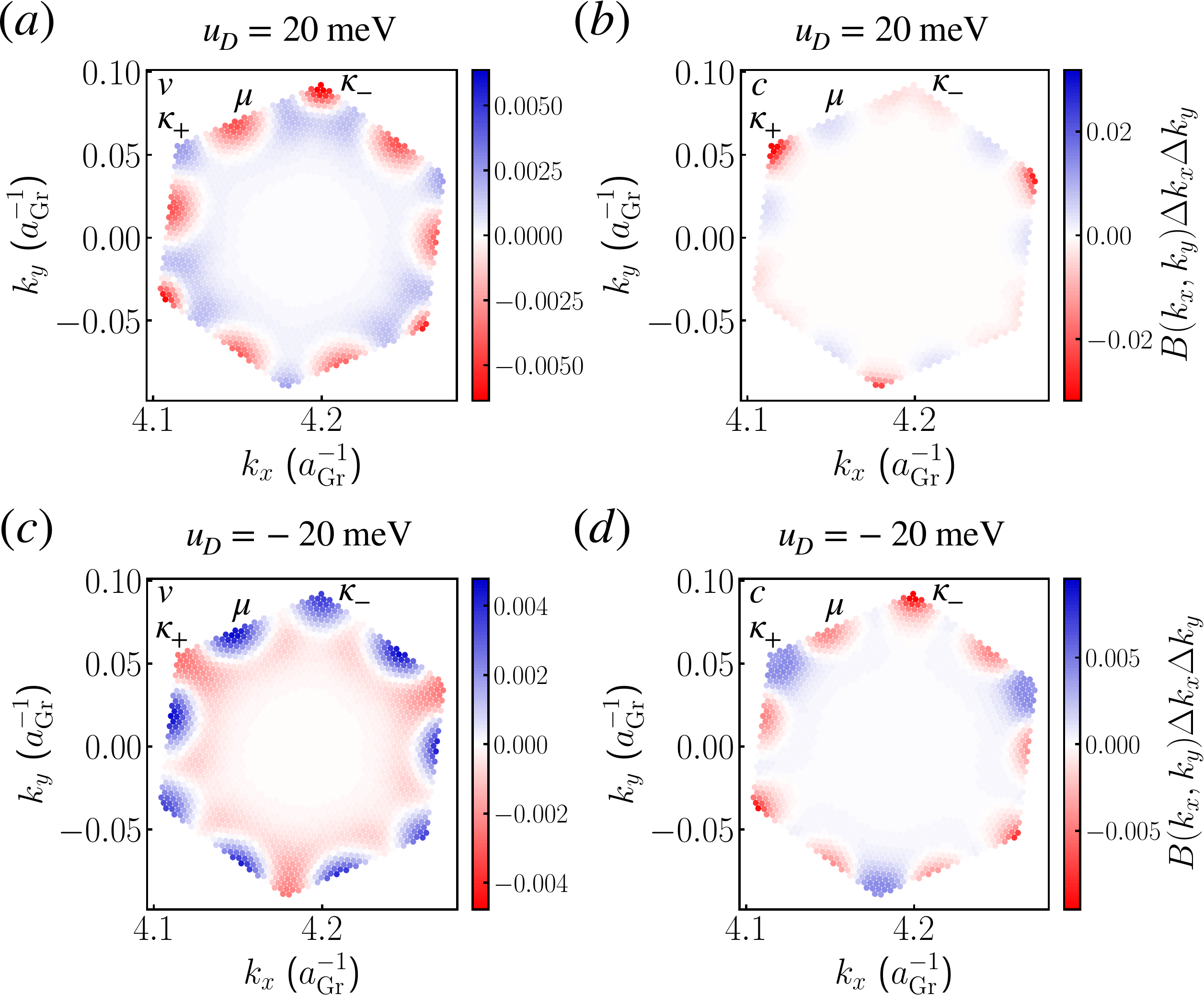}
    \caption{\emph{Electronic} Berry curvature for the occupied ($v$) and empty band ($c$) for parameters $\theta = 0.77^\circ$ and $u_D = \pm 20\:\mathrm{meV}$.}
    \label{fig:nonInteractingBerryCurvature}
\end{figure}

Using the model outlined above, we calculate the electronic dispersion of the hBN/R5G/hBN stack. Examples of the non-interacting band structures for $\theta = 0.77^\circ$ and $u_D = \pm 20 \:\mathrm{meV}$ are shown in Fig.~\ref{fig:nonIntDispersion}. The hBN layers fold the bands of isolated rhombohedral graphene into the moiré Brillouin zone (mBZ) and hybridize the folded bands, resulting in two gapped bands immediately above (labeled $c$) and below the Fermi level (labeled $v$) at charge neutrality. The hBN moiré potential preserves $C_3$ symmetry, allowing us to use symmetry indicators to compute the Chern number (mod 3) of the electronic bands. The $C_3$ eigenvalues can take the values $1$, $\omega$, or $\omega^*$, where $\omega = e^{\mathrm{i} \frac{2\pi}{3}}$. The Chern number $C$ is related to the product of the $C_3$ eigenvalues ($\lambda^{C_3}_{k}$) at the high-symmetry points as follows:~\cite{CnChernNumber}
\begin{equation} e^{\mathrm{i} \frac{2\pi}{3} C} = \prod_{k \in \{\gamma, \kappa_+, \kappa_-\}} \lambda^{C_3}_{k}.
\end{equation}
For $u_D = 20\: \mathrm{meV}$, we find that the empty (conduction) band has a Chern number of 1, while the occupied (valence) band has a Chern number of 0. For $u_D = -20\: \mathrm{meV}$, both bands have Chern number 0. In Appendix~\ref{Apx:PhaseDiagramofNonIntModel} we show that a variety of possible Chern numbers for both bands can be achieved in this model by tuning the displacement field $u_D$ and twist angle $\theta$. In addition, we show explicitly how tuning the moire potential gaps out the bands folded to the mBZ to generate a band with Chern number 1 at $u_D = 20\: \mathrm{meV}$. Figure~\ref{fig:nonInteractingBerryCurvature} shows the Berry curvature of the non-interacting bands for the chosen parameters, revealing that it is concentrated at the edges of the mBZ. At $u_D = 20 \:\mathrm{meV}$ the maxima and minima of the bands are also at the edge of the mBZ, indicating that low-energy charge-neutral excitations are dominated by states near the mBZ edges. We therefore expect this large electronic Berry curvature to also be visible in the exciton Berry curvature.

\section{Excitons}
\label{Sec:Excitons}
We next study the exciton bound states of hBN/R5G/hBN, their dispersion and Berry curvature.

\subsection{Methods}
We calculate the exciton dispersion and wave functions in hBN/R5G/hBN stacks that form from the two well separated bands shown Fig.~\ref{fig:nonIntDispersion}. At charge neutrality the ground state has the lower band ($v$) occupied and the upper band ($c$) unoccupied. We will consider excitons formed from these two bands where $c_{k, c}$ ($c_{k, v}$) annihilates an electron in the conduction (valence) band at crystal momentum $k$. We restrict our calculation to just these two bands since we are interested in the low energy spectrum which will not be effected significantly by the inclusion of higher and lower energy bands. The wave function of an exciton at momentum $\boldsymbol{p}$ is expressed as: 
\begin{equation} \ket{\psi^{\boldsymbol{p}}} = \sum_{\boldsymbol{k}\in \mathrm{mBZ}}\phi^{\boldsymbol{p}}_{\boldsymbol{k}} c^\dagger_{\boldsymbol{p}+\boldsymbol{k}, c} c_{\boldsymbol{k}, v} \ket{\mathrm{GS}}, \label{Eq:ExcitonWavefunction} \end{equation}
where $\ket{\mathrm{GS}}$ denotes the non-interacting ground state, projected onto the two bands of interest (e.g., $\ket{\mathrm{GS}} = \prod_{\boldsymbol{k}} c^\dagger_{\boldsymbol{k}, v} \ket{0}$). We study the spin-valley polarized regime and so we restrict our focus to excitons at just one valley and a single spin species~\cite{VishwanathPaper}.

To model the excitons we project both the kinetic and interaction Hamiltonians into the basis $c^\dagger_{\boldsymbol{p}+\boldsymbol{k}, c} c_{\boldsymbol{k}, v} \ket{\mathrm{GS}}$~\cite{TrionsSchindler, KhalafSkyrmions}. The resulting exciton Hamiltonian, detailed in Appendix~\ref{Apx:ExcitonHamiltonian}, is written in terms of non-interacting single-particle energies and the products of the non-interacting electronic wave functions (the ``form factors"). These products are given by:
\begin{equation} U^{\boldsymbol{p}, \boldsymbol{k}, \boldsymbol{k}'}_{\alpha, \alpha', \beta, \beta'} = \sum_{\boldsymbol{Q}} V(\boldsymbol{p}+\boldsymbol{Q}) \braket{u_{\boldsymbol{p}+\boldsymbol{k}+\boldsymbol{Q}}^\alpha|u_{\boldsymbol{p}}^{\alpha'}}\braket{u_{\boldsymbol{p}'}^{\beta}|u_{\boldsymbol{p}+\boldsymbol{k}'+\boldsymbol{Q}}^{\beta'}},
\label{eq:BlochProducts} \end{equation}
where $\ket{u_{\boldsymbol{p}}^{\alpha}}$ represents the eigenvector of the moiré Hamiltonian [Eq.~\eqref{eq:MoireHamiltonian}] for momentum $\boldsymbol{p}$ and band $\alpha$. We use the double-gated Coulomb potential, which accounts for the screening effects introduced by the metal electrodes either side of the sample. This potential is given by,
\begin{equation} V(|\boldsymbol{p}|) = \frac{V_0 \tanh(|\boldsymbol{p}|d)}{|\boldsymbol{p}|}, \end{equation}
where $d$ is the metal electrode gate distance. We use $d = 250\: \text{\r{A}}$~\cite{VishwanathPaper} for the following calculations. The interaction strength $V_0 = \frac{2\pi}{\epsilon_0 \epsilon_r}$ depends on the relative permittivity $\epsilon_r$, we use $\epsilon_r = 6$~\cite{bernevig2025berrytrashcanmodelinteracting}. We calculate the exciton Berry curvature using the gauge invariant method of Ref.~\onlinecite{discreteBZ_ChernNumbers, davenport2025excitonberryology} (see Appendix~\ref{Apx:ExcitonBerryCurvature} for details).

\subsection{Exciton dispersion and Berry curvature}
\begin{figure}
    \centering
    \includegraphics[width=1.0\linewidth]{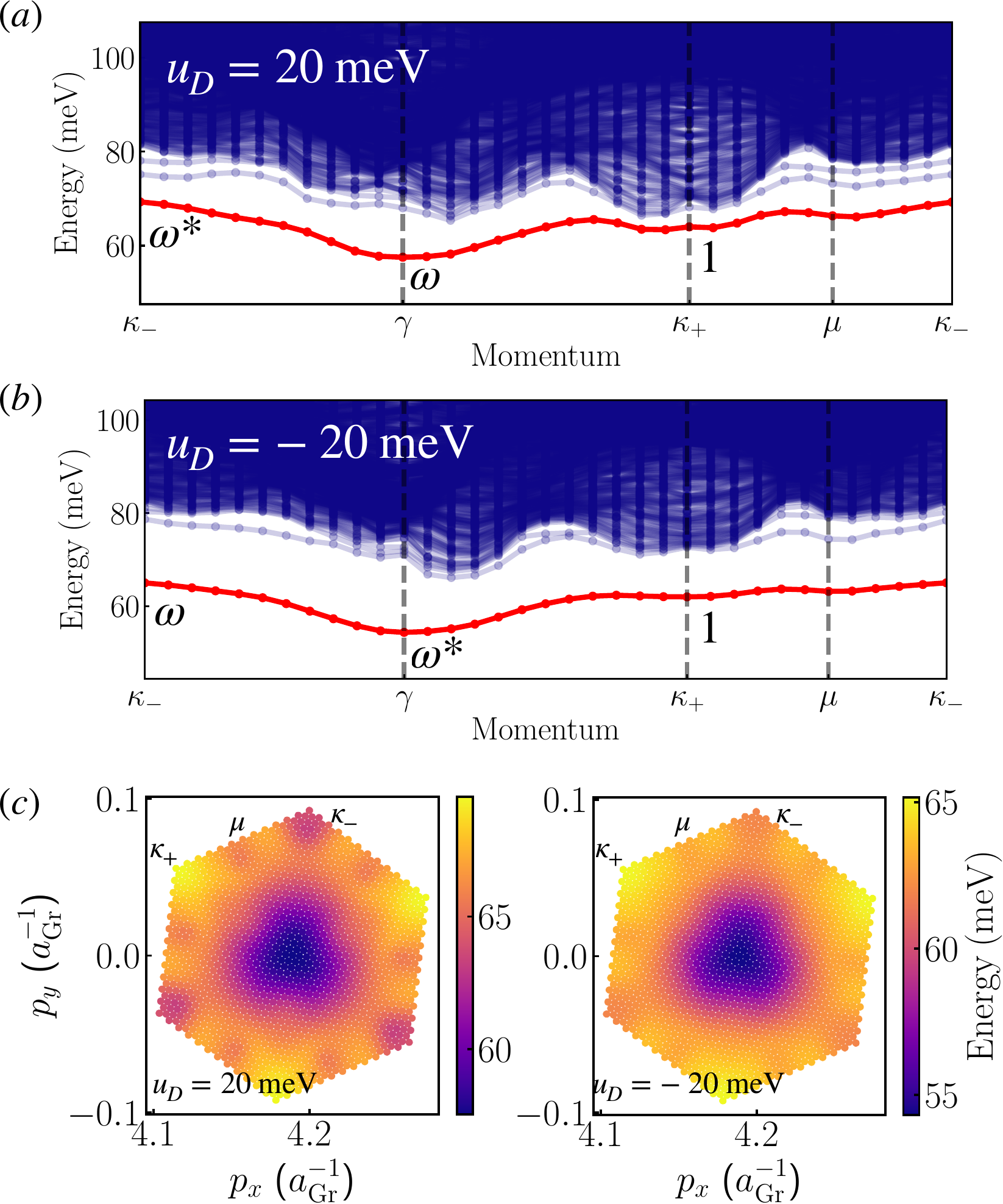}
    \caption{Exciton dispersion (in red) with particle-hole continuum above (in blue) for $\theta = 0.77^\circ$ $(a)$ $u_D = 20 \:\mathrm{meV}$ and $(b)$ $u_D = -20 \:\mathrm{meV}$. In both cases the $C_3$ symmetry eigenvalues of the excitons are marked. The dispersion of the excitons throughout the whole BZ for both parameter choices is shown in $(c)$.}
    \label{fig:ExcitonDispersion}
\end{figure}
\begin{table}[h]
    \centering
    \begin{tabular}{c|cccc}
        \hline
        $u_D$ & $\lambda^{\mathrm{exc}}_{\gamma}$ & $\lambda^{\mathrm{exc}}_{\kappa_+}$ & $\lambda^{\mathrm{exc}}_{\kappa_-}$ & $\boldsymbol{s}_{\mathrm{exc}}$\\
        \hline
        $20$ meV & $\omega$ & $1$ & $\omega^*$ & $1c$\\
        $-20$ meV & $\omega^*$ & $1$ & $\omega$ & $1b$ \\
        \hline
    \end{tabular}
    \caption{\( C_3 \) eigenvalues of the exciton wave function at high-symmetry points in the Brillouin zone for two different parameter choices. The exciton Wannier centre $\boldsymbol{s}_{\mathrm{exc}}$ is stated in terms of the Wyckoff positions (see Fig.~\ref{fig:mBZandUnitCell}b).}
    \label{tab:C3eigenvalues}
\end{table}

Fig.~\ref{fig:ExcitonDispersion} shows the exciton dispersions at a twist angle of $\theta = 0.77^\circ$ under a displacement field of $u_D = 20\:\mathrm{meV}$ and $u_D = -20 \:\mathrm{meV}$. For both parameter choices, the exciton binding energy is largest around the $\gamma$ point since the interaction potential $V(p)$ peaks at $p = 0$. The excitons in both cases are fully gapped across the mBZ but the maximum binding energy for $u_D = -20 \:\mathrm{meV}$ is substantially larger at $\sim 17\:\mathrm{meV}$ compared to $\sim11\:\mathrm{meV}$ for $u_D = 20 \:\mathrm{meV}$. We can use symmetry indicators to study the exciton band topology. We note that in Ref.~\cite{davenport2025excitonberryology} we found that there were an infinite family of exciton Berry connections which give different (but related) exciton Berry curvatures. However, we showed that, despite this, the exciton Chern number is still uniquely defined. We compute the total $C_3$ eigenvalues of the many body exciton wave function at the high-symmetry points in the Brillouin zone (see  Fig.~\ref{fig:ExcitonDispersion}). In both parameter regimes the exciton Chern number of the band is $0$. However, despite the trivial Chern number, there is interesting \emph{crystalline} topology. The symmetry indicators of the excitons can be used to calculate the centre of the maximally localised exciton Wannier states for the exciton band~\cite{jonahMLXWF, davenport2025excitonberryology, ExcitonTopologyDavenport}. In Tab.~\ref{tab:C3eigenvalues} we show the shift of the maximally localised exciton Wannier function from the centre of the moiré unit cell $\boldsymbol{s}_{\mathrm{exc}}$ (see Appendix~\ref{Apx:ExcitonWannerStates} for details of this calculation). In both cases, the exciton Wannier centre is not at the centre of the moiré unit cell. In addition, the exciton Wannier centre varies between the two configurations: at $u_D = 20\:\mathrm{meV}$ it sits at the $1c$ Wyckoff position but for $u_D = -20\:\mathrm{meV}$ it is at the $1b$ Wyckoff positions.

\begin{figure*}[]
    \centering
    \includegraphics[width=0.98\textwidth]{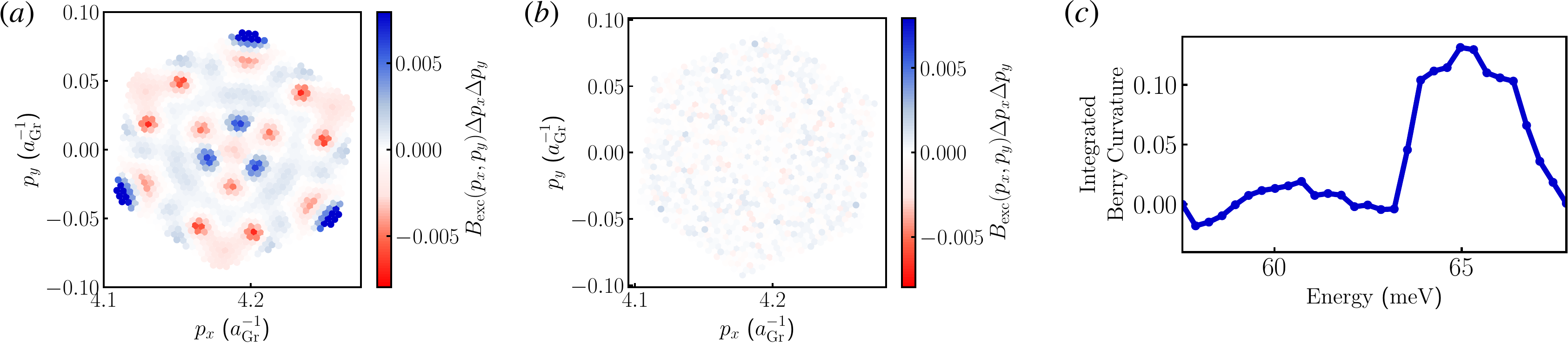}
    \caption{Exciton Berry curvature for the lowest energy exciton band for $\theta = 0.77^\circ$, $(a):$ $u_D = 20 \:\mathrm{meV}$ and $(b):$ $u_D = -20 \:\mathrm{meV}$. Panel $(c)$ is the exciton Berry curvature integrated up to the given energy cutoff for $u_D = 20\: \mathrm{meV}$ (not shown for full mBZ).}
    \label{fig:ExcitonBerryCurvature}
\end{figure*}
Beyond the global topology of the exciton band, the exciton Berry curvature (Fig.~\ref{fig:ExcitonBerryCurvature}) reveals marked differences between the two displacement field values. For $u_D = 20\:\mathrm{meV}$, the curvature exhibits pronounced minima and maxima, despite the total Chern number of the exciton band being zero. These features originate from the underlying \emph{electronic} Berry curvature. Specifically, the enhanced \emph{exciton} Berry curvature near the $\gamma$ point can be seen to be inherited from the \emph{electronic} Berry curvature. In Fig.~\ref{fig:nonIntDispersion}a we can see that the lowest energy particle hole excitations take an electron from the $\kappa_+$ point in the lower band to $\mu$ in the upper band. Since the electronic Berry curvature is strongly concentrated at $\kappa_+$ (Fig.~\ref{fig:nonInteractingBerryCurvature}a), excitons with momentum $\boldsymbol{p} \approx \boldsymbol{\mu} - \boldsymbol{\kappa_+}$ (or $C_3$ rotations of this momentum) inherit this curvature. In contrast, for $u_D = -20\:\mathrm{meV}$, the exciton Berry curvature is nearly uniform and vanishing (Fig.~\ref{fig:ExcitonBerryCurvature}b). This occurs because the electronic bands do not have as pronounced maxima at the edge of the mBZ compared to the $u_D = 20\: \mathrm{meV}$ case. Therefore the exciton bands are mainly composed of electronic states from the centre of the mBZ. The Berry curvature of the electronic bands at the centre can be seen to be very close to 0 (see Fig.~\ref{fig:nonInteractingBerryCurvature}c and~\ref{fig:nonInteractingBerryCurvature}d) and so the exciton Berry curvature is similarly small in magnitude.

Excitons will thermally occupy the exciton band leading to an enhanced population at the band minimum. Therefore, whilst the overall exciton band is topologically trivial, the locally enhanced Berry curvature in the vicinity of the band minimum (at $u_D = 20\: \mathrm{meV}$) will influence exciton transport behaviour. The integrated exciton Berry curvature, as a function of the energy cutoff, is shown in Fig.~\ref{fig:ExcitonBerryCurvature}c. Due to the presence of both minima and maxima in the Berry curvature, the integrated value oscillates for $u_D = 20\: \mathrm{meV}$. This will lead to a temperature dependence of response functions sensitive to the exciton Berry curvature, as thermal excitation allows excitons to occupy states beyond the band minimum, altering the Boltzmann-weighted Berry curvature.  In contrast at $u_D = -20 \:\mathrm{meV}$, the uniform (and zero) Berry curvature of the exciton band means that there will not be temperature dependent exciton Berry curvature effects. Hence, rhombohedral graphene provides a platform in which it is possible to electrically tune between regimes with and without \emph{excitonic} Berry curvature effects.

\section{Discussion}
\label{sec:Discussion}
The response and transport properties of electrons in non-interacting bands are well known to be influenced by the electronic Berry curvature~\cite{BerryPhaseEffects}. Fully elucidating the transport effects of exciton Berry curvature represents a new frontier in the study of quantum transport phenomena.  Our work implies that rhombohedral graphene is a possibly important material for exploring the interplay of excitons and topology. For $u_D = -20\:\mathrm{meV}$, the Berry curvature of the single electron dispersion of rhombohedral graphene is concentrated at the edge of the mBZ, exactly where the low energy excitons are formed, meaning that the lowest energy excitons in rhombohedral graphene inherit the Berry curvature of the electronic bands. The transport properties of excitons are likely to be influenced by this exciton Berry curvature. One possible measurable consequence of the enhanced Berry curvature at a displacement field of $u_D = 20\: \mathrm{meV}$ is a thermal Hall effect~\cite{zhang2024thermal}. Applying an in-plane temperature gradient to the sample would generate a transverse exciton current, \emph{i.e.} most clearly visible as a chiral exciton current at the edge of the system. Typically such exciton currents are hard to measure because excitons are neutral,  however, as we have seen, the conduction band and valence band electronic wave functions are layer polarised. Hence, the exciton flow could be detected by measuring the electron or hole current on either the top or bottom layer of the sample similar to a drag current~\cite{nguyen2309perfect,qi2025perfect}. We note that in general the thermal hall effect is suppressed by the gap from the ground state to the excitation band. Despite this the thermal Hall effect can still be measurable experimentally \emph{e.g.} for topological magnons~\cite{thermalHallMeasurementMagnons}. However, observing the thermal Hall effect with excitons may in fact in general be easier than for magnons because excitons are easy to pump unlike magnons.

We have also predicted that the maximally localised exciton Wannier states are shifted away from the centre of the unit cell, directly analogous to the concept of an obstructed atomic insulator in non-interacting electron systems~\cite{TQCPaper}. Obstructed atomic insulators are known to possess crystal defect responses that depend on the Wannier state shift, as well as corner or edge polarization~\cite{defects1}. Hence we expect that the exciton Wannier state shift in rhombohedral graphene could analogously lead to new excitonic crystal defect responses and exciton corner or edge modes~\cite{ExcitonTopologyDavenport}. The current spatial resolution of exciton probes would not allow it to be possible to detect these effects in bare graphene with a unit cell of order $2.46 \:\text{\r{A}}$, but the enhanced moiré unit cell size increases the shift to within the resolution range of techniques like EELs, which can achieve spatial resolutions on the order of $\sim \mathrm{nm}$~\cite{ExcitonImage1, ExcitonImage2}. Moiré systems such as hBN/R5G/hBN therefore offer a promising platform for experimentally probing exciton crystalline topology. However, understanding the full range of experimental signatures, beyond edge states~\cite{ExcitonTopologyDavenport}, remains a compelling and open direction for future research.
\\
\begin{acknowledgments}
We would like to thank Johannes Lischner for useful discussions. We acknowledge support from the Imperial-TUM flagship partnership. This work was supported by a UKRI Future Leaders Fellowship MR/Y017331/1. HD acknowledges support from the Engineering and Physical Sciences Research Council (grant number EP/W524323/1). JK acknowledges support from the Deutsche Forschungsgemeinschaft (DFG, German Research Foundation) under Germany’s Excellence Strategy–EXC– 2111–390814868, DFG grants No. KN1254/1-2, KN1254/2- 1, and TRR 360 - 492547816; as well as the Munich Quantum Valley, which is supported by the Bavarian state government with funds from the Hightech Agenda Bayern Plus.
\end{acknowledgments}

\section*{Data Availability}
The data that support the findings of this article are openly available~\cite{davenport_2025_16902666}.

\bibliography{refs}

\newpage
\onecolumngrid
\appendix
\section{Pentalayer rhombohedral graphene}
\label{apdx:FullModelAndLowEnergyModel}
We derive the low energy model for pentalayer rhombohedral graphene (R5G) introduced in the main text. The model is derived from the  full 10 band model used by Ref.~\onlinecite{VishwanathPaper}. We begin by summarising this full 10 band model before introducing the low energy 2 band model we use for the remainder of the paper. 
\subsection{Full pentalayer model}
We describe the full model of R5G, see Ref.~\onlinecite{VishwanathPaper} for more details. Graphene has a lattice parameter $a = 2.46 \mathrm{\r{A}}$ and lattice vectors,
\begin{align}
\boldsymbol{R}_1 &= \left(a, 0\right)\\
\boldsymbol{R}_2 &= a\left(\frac{1}{2}, \frac{\sqrt{3}}{2}\right).
\end{align}
The corresponding reciprocal lattice vectors are,
\begin{align}
\boldsymbol{G}_1 &= \frac{2\pi}{a} \left(1, -\frac{1}{\sqrt{3}}\right)\\
\boldsymbol{G}_2 &= \frac{2\pi}{a} \left(0, \frac{2}{\sqrt{3}}\right).
\end{align}
The high symmetry points of graphene are,
\begin{align}
\boldsymbol{K}_{Gr} &= \frac{2}{3} \boldsymbol{G}_1 + \frac{1}{3} \boldsymbol{G}_2\\
\boldsymbol{K}'_{Gr} &= \frac{1}{3} \boldsymbol{G}_1 + \frac{2}{3} \boldsymbol{G}_2.
\end{align}
We can write the Hamiltonian for R5G in layer space as,
\begin{equation}
h_{R5G}(\boldsymbol{k}) = \begin{bmatrix}
h_{0}^{(0)} & h^{(1)} & h^{(2)} &0 & 0 \\
[h^{(1)}]^\dagger & h_{1}^{(0)} & h^{(1)} & h^{(2)} & 0 \\
[h^{(2)}]^\dagger & [h^{(1)}]^\dagger & h_{2}^{(0)} &  h^{(1)} & h^{(2)} \\
0 & [h^{(2)}]^\dagger & [h^{(1)}]^\dagger & h_{3}^{(0)} &  h^{(1)} \\
0 & 0 & [h^{(2)}]^\dagger & [h^{(1)}]^\dagger & h_{4}^{(0)},
\end{bmatrix}
\end{equation}
where the rows and columns correspond to different layers of graphene. The $h^{(0)}_{l}$, $h^{(1)}$ and $h^{(2)}$ matrices are written in the basis of the graphene sublattice orbitals for each layer ($l$) of graphene in the pentalayer stack. These matrices are,
\begin{align}
h^{(0)}_{l}(\boldsymbol{k}) &= \begin{pmatrix} u_{l} & -t_0 f_{\boldsymbol{k}} \\ -t_0 \bar f_{\boldsymbol{k}} & u_{l}\end{pmatrix},\\
h^{(1)}(\boldsymbol{k}) &= \begin{pmatrix} t_4 & t_3 \bar f_{\boldsymbol{k}} \\ t_1 \bar f_{\boldsymbol{k}} & t_4 f_{\boldsymbol{k}}\end{pmatrix},\\
h^{(2)}(\boldsymbol{k}) &= \begin{pmatrix} 0 & \frac{t_2}{2} \\ 0  & 0\end{pmatrix}.
\end{align}
We define,
\begin{equation}
f_{\boldsymbol{k}} = \sum_{i = 0}^2 e^{\mathrm{i} \boldsymbol{k}.\boldsymbol{\delta}_i},
\end{equation}
with $\boldsymbol{\delta}_n$ defined as the vectors between nearest neighbour carbon atoms in each layer~\cite{VishwanathPaper}. We use the hopping values $t_0 = 3.1 \: \mathrm{eV}$, $t_1 = 380\: \mathrm{meV}$, $t_2 = -21\:\mathrm{meV}$, $t_3 = 290\:\mathrm{meV}$ and $t_4 = 141\:\mathrm{meV}$~\cite{VishwanathPaper, MoirePotential}. The applied electric field gives a layer dependent electric potential,
\begin{equation}
u_{l} = u_D l
\end{equation}
with the layer index $l$ and the tunable displacement field $u_D$. 
\subsection{Low energy model}
We now derive the low energy model of pentalayer rhombohedral graphene. For finite $u_D$ there are two bands immediately above and below the Fermi energy at charge neutrality. These two bands are layer polarised around the $\boldsymbol{K}_{Gr}$ point. The band which is layer polarised to layer $1$ has most of the wave function weight on the $A$ sublattice whereas the layer $5$ polarised band is on the $B$ sublattice. The low energy physics, around $\boldsymbol{K}_{Gr}$, can therefore be well described by a model solely in the subspace $1A$, $5B$ (where we label the positions in the unit cell by the layer $1, 5$ and the sublattice $A, B$). We rewrite the equation for the energy dispersion $\epsilon(\boldsymbol{k})$ as,
\begin{equation}
\begin{pmatrix}
    h(\boldsymbol{k})_{11} & h(\boldsymbol{k})_{12}\\
     h(\boldsymbol{k})_{21} &  h(\boldsymbol{k})_{22}
\end{pmatrix} \begin{pmatrix}
    \boldsymbol{c}_{1}(\boldsymbol{k})\\ \boldsymbol{c}_{2}(\boldsymbol{k})
\end{pmatrix} = \epsilon(\boldsymbol{k}) \begin{pmatrix}
    \boldsymbol{c}_{1}(\boldsymbol{k})\\ \boldsymbol{c}_{2}(\boldsymbol{k})
\end{pmatrix}\label{eq:fullHam},
\end{equation}
where the matrix $h(\boldsymbol{k})_{11}$ represents the  block of the full matrix $h(\boldsymbol{k})$ in  the subspace $1A, 5B$, $h(\boldsymbol{k})_{22}$ is the high energy subspace (\emph{i.e.} $2B, \dots 5A$), and $h(\boldsymbol{k})_{12}$ and $h(\boldsymbol{k})_{21}$ represents the coupling between the subspaces. The components of the Bloch function in each subspace are given by the functions $\boldsymbol{c}_{1}(\boldsymbol{k})$ and $\boldsymbol{c}_{2}(\boldsymbol{k})$. We wish to find a $2\times2$ effective Hamiltonian in the low energy subspace~\cite{MacdonaldGreensFunctionMethod}. To do this we expand Eq.~\eqref{eq:fullHam} to give two relations,
\begin{align}
\boldsymbol{c}_{1}(\boldsymbol{k}) &=(\epsilon - h(\boldsymbol{k})_{11})^{-1} h(\boldsymbol{k})_{12}  \boldsymbol{c}_{2}(\boldsymbol{k})\\ \boldsymbol{c}_{2}(\boldsymbol{k}) &= (\epsilon - h(\boldsymbol{k})_{22})^{-1} h(\boldsymbol{k})_{21}  \boldsymbol{c}_{1}(\boldsymbol{k}).
\end{align}
Combining these we obtain an equation for the wave function components in the low energy subspace $\boldsymbol{c}_{1}(\boldsymbol{k})$,
\begin{equation}
[h_{11}(\boldsymbol{k}) + h_{12}(\boldsymbol{k}) (h_{22}(\boldsymbol{k}) - \epsilon)^{-1} h_{21}(\boldsymbol{k}) - \epsilon] \boldsymbol{c}_{1}(\boldsymbol{k}) =0.
\end{equation}
We expand to first order in energy $\epsilon$ and obtain the effective Hamiltonian,~\cite{MacdonaldGreensFunctionMethod}
\begin{equation}
\begin{aligned}
h_{\mathrm{eff}}(\boldsymbol{k}) = &\big\{1-h(\boldsymbol{k})_{12} \big[h(\boldsymbol{k})_{22}\big]^{-2} h(\boldsymbol{k})_{21}\big\}^{-1} \\&\big\{h(\boldsymbol{k})_{11} + h(\boldsymbol{k})_{12} \big[h(\boldsymbol{k})_{22}\big]^{-1} h(\boldsymbol{k})_{21}\big\}.
\end{aligned}
\end{equation}
We consider small momenta around the $\boldsymbol{K}_{Gr}$ valley of the R5G, and so we Taylor expand $h_{\mathrm{eff}}(\boldsymbol{k})$ in powers of $\boldsymbol{k}$ about $\boldsymbol{K}_{Gr}$. For pentalayer rhombohedral graphene, the dominant off-diagonal terms in $h_{\mathrm{eff}}(\boldsymbol{k})$ go as $\sim |\boldsymbol{k}|^5$ (as described in Ref.~\cite{SenthilPaper}). When we expand the diagonal elements to zeroth order in $\boldsymbol{k}$ we also obtain a contribution from the applied displacement field $u_D$. We introduce a further correction beyond the model described in Ref.~\onlinecite{SenthilPaper}; the full model has a characteristic quadratic curvature which dominates at small momenta $\boldsymbol{k}$ from the Dirac point $\boldsymbol{K}_{Gr}$. We find that this effect can be captured in the low energy model by expanding the diagonal components to second order in momentum $\boldsymbol{k}$. This gives an effective hamiltonian in the low energy subspace which is,
\begin{equation}
\begin{aligned}
    &h_{\mathrm{eff}}(k_x, k_y) = \\ &\left(
\begin{array}{cc}
 2 u_D+\gamma_-(k_x^2+k_y^2) & \frac{v_F^5}{t_0^4}\left(k_x-i k_y\right)^5 \\
 \frac{v_F^5}{t_0^4}\left(k_x+i k_y\right)^5 & -2 u_D+\gamma_+(k_x^2+k_y^2) \\
\end{array}
\right).
\label{eq:2by2Model}
\end{aligned}
\end{equation}
The $\gamma_{0/1}$ take a simple form if we assume small $u_D$ (specifically $u_D < v_3/a \ll v_F/a$) and expand to first order in $u_D$,
\begin{align}
\gamma_\pm &\approx \frac{2 v_F  v_4}{t_1}\pm\frac{u_D v_F^2}{t_1^2}.
\label{eq:gammaCorrections}
\end{align}
This simplified model predicts a sign change in the curvature of the higher-energy band at the $\boldsymbol{K}_{Gr}$ point at a displacement field of \( u_D = \frac{2v_4 t_1}{v_F} \sim 35 \:\mathrm{meV} \), while the band dispersion curvature at the $\boldsymbol{K}_{Gr}$ point of the lower-energy band remains unchanged for all \( u_D \geq 0 \). This behaviour in good agreement with results from the full model. Note that this curvature sign change is not captured in the simple model described in Ref.~\onlinecite{SenthilPaper} so this model more accurately describes the low energy physics of R5G. A comparison between the dispersions of the full and simplified models for a representative parameter set is shown in the main text. As expected, the two models agree well at momenta close to $\boldsymbol{K}_{Gr}$, with increasing deviation at larger momenta. Notably, the full model lacks inversion symmetry about the $\boldsymbol{K}_{Gr}$ point—i.e., it is not symmetric under \( (k_x, k_y) \rightarrow (-k_x, -k_y) \)—unlike the low-energy approximation. This asymmetry arises from the inclusion of intra-layer hopping \( t_3 \). When \( t_3 \) is neglected, the system possesses a unitary symmetry \( U = \bigoplus_{l=0}^{N_l-1}\,(-1)^l \sigma_z \), obeying \( U h(\boldsymbol{k}) U^\dagger = h(-\boldsymbol{k}) \). This makes the spectrum symmetric about about the $\boldsymbol{K}_{Gr}$. Since the low-energy \( 2 \times 2 \) model does not include \( t_3 \), it is symmetric about the $\boldsymbol{K}_{Gr}$ point. In addition to capturing the low energy dispersion, the low energy model derived above also captures the Berry curvature distribution of the full model (as the model without the quadratic correction does too~\cite{SenthilPaper}).

\section{Moiré Hamiltonian}
\label{apx:MoireHamiltonian}
We study R5G between hBN layers \emph{i.e.} hBN/R5G/hBN stacks. We use the same model as Ref.~\onlinecite{VishwanathPaper}. The lattice mismatch between the hBN and the R5G leads to a moiré pattern. There are two sources of the lattice mismatch: firstly there is a lattice constant mismatch between hBN and graphene, $\epsilon = a_{\mathrm{hBN}}/{a_{Gr}} = 0.018$. Secondly we can rotate the hBN by some angle $\theta$ to change the moiré periodicity. We define a rotation matrix $R_\theta$ to describe rotation of the hBN graphene by angle $\theta$ with respect to the R5G. Similarly the matrix $M = (1+\epsilon) I$ can be used to calculate the effects of the lattice constant mismatch. The moiré reciprocal lattice vectors are,
\begin{equation}
\boldsymbol{g}_j = [I - M^{-1} R_{\theta}] \boldsymbol{G}_j,
\end{equation}
for $j\in \{1, 2\}$~\cite{VishwanathPaper, MoirePotential}. In addition we define,
\begin{equation}
\boldsymbol{g}_3 = \boldsymbol{g}_1 + \boldsymbol{g}_2.
\end{equation}
The high symmetry points in the new moiré Brillouin zone are,
\begin{align}
\boldsymbol{\gamma} &= \boldsymbol{K}_{\mathrm{Gr}}\\
\boldsymbol{\kappa}^+ &= \boldsymbol{\gamma} - \frac{1}{3} \boldsymbol{g}_1 + \frac{1}{3} \boldsymbol{g}_2\\
\boldsymbol{\kappa}^- &= \boldsymbol{\gamma} - \frac{2}{3} \boldsymbol{g}_1 - \frac{1}{3} \boldsymbol{g}_2
\\
\boldsymbol{\mu} &= \boldsymbol{\gamma} - \frac{1}{2} \boldsymbol{g}_1.
\end{align}

We model the effect of the hBN layers as an effective potential acting on the outermost layers ($l=1, 5$) of the R5G. 
This moiré potential can be written as,
\begin{equation}
H_{V} = \sum_{\substack{\boldsymbol r\\ l\in \{\mathrm{top}, \mathrm{btm}\}}} V_{\mathrm{eff}}^{l}(\boldsymbol r) c^\dagger_{\boldsymbol r, l} c_{\boldsymbol r, l},
\end{equation}
$\boldsymbol r$ goes over all unit cells and $c^\dagger_{\boldsymbol r, l}$ creates an electron at position $\bs{r}$ and layer $l$. 
Since our effective Hamiltonian retains only the $A$ sublattice in the bottom layer and the $B$ sublattice in the top layer, we apply the moiré potential exclusively to these sites. The moiré potential has the moiré periodicity so we can write the potential as,
\begin{equation}
V^{l}_{\mathrm{eff}}(\boldsymbol{r}) = \sum_{\substack{\eta \in \{1, -1\} \\ j\in \{1, 2, 3\}}} V_{l, \eta\boldsymbol{g}_j} \exp\left[\mathrm{i}\eta\boldsymbol{g}_j \cdot \boldsymbol{r}\right]
\end{equation}
Here, the sum runs over the first shell of moiré reciprocal lattice vectors around the first moiré BZ. This is known as the ``first harmonic” approximation. We use values for the prefactors $V_{l, \eta\boldsymbol{g}_j}$ introduced in Ref.~\cite{MoirePotential} such that the potential can be expressed as,
\begin{equation} V^{l}_{\mathrm{eff}}(\boldsymbol{r}) = V_0 +  V_1 \sum_{\substack{\eta \in \{1, -1\} \\ j\in \{1, 2, 3\}}}
\exp\left[\mathrm{i}\eta\left(\boldsymbol{g}_j \cdot \boldsymbol{r} + \psi_{\xi}+\phi_l \right)\right]. 
\end{equation} 
There is a phase shift $\psi_{\xi}$ which depends on the the stacking configuration, for example, the $A$ sublattice of graphene may lie above either a boron atom ($\xi = 0$) or a nitrogen atom ($\xi = 1$), as described in Ref.~\cite{MoirePotential} and this leads to differing phases for the effective potential. We use the values in Refs.~\cite{MoirePotential} and restrict our study to the $\xi = 0$ stacking for both the top and bottom layers, for which $\psi_{\xi} = 223.5^\circ$. The additional phase factor $\phi_l$ is 0 for the bottom layer but for $N_l$ layers of graphene, the top layer gains a phase $\phi_l = \frac{2\pi N_l}{3}$. This phase accounts for the relative shift in the moiré pattern that arises because, in a given unit cell, the $B$ sublattice site in the top layer is not directly aligned above the $A$ sublattice sites in the bottom layer~\cite{BernevigRhombohedral}. Note that the moiré coupling does not couple the top and bottom layers. We use $V_0 = 28.9 \:\mathrm{meV}$ and $V_1 = 21.0\:\mathrm{meV}$ following Refs.~\onlinecite{MoirePotential, VishwanathPaper}.

We rewrite the Hamiltonian in momentum space using, 
\begin{equation}
    c^\dagger_{\boldsymbol{r}, l} = \frac{1}{L} \sum_{\boldsymbol{k}} e^{i\boldsymbol{k}\cdot\boldsymbol{r}} c^\dagger_{\boldsymbol{k}, l}.
\end{equation}
where the sum over $\boldsymbol{k}$ goes over all momenta in the R5G BZ and the system consists of $L^2$ unit cells. This gives,
\begin{equation}
H_V = \sum_{l, \boldsymbol{r}, \boldsymbol{k}, \boldsymbol{k}'} \sum_{\substack{\eta \in \{1, -1\} \\ j\in \{1, 2, 3\}}} e^{i(\boldsymbol{k} - \boldsymbol{k}' + \boldsymbol{g}_j) \cdot \boldsymbol{r}} V_{l, \eta\boldsymbol{g}_j} c^\dagger_{\boldsymbol{k}, l} c_{\boldsymbol{k}', l}.
\end{equation}
Performing the summation over $\boldsymbol{r}$ gives,
\begin{equation}
H_V = \sum_{l, \boldsymbol{k}, \boldsymbol{k}'} \sum_{\substack{\eta \in \{1, -1\} \\ j\in \{1, 2, 3\}}} \delta_{\boldsymbol{k}, \boldsymbol{k}' - \eta\boldsymbol{g}_j} V_{l, \eta\boldsymbol{g}_j} c^\dagger_{\boldsymbol{k}, l} c_{\boldsymbol{k}', l}.
\end{equation}
We now express all momenta in terms of the momentum in the mBZ ($\boldsymbol{k}$) and a moiré reciprocal lattice vector ($\boldsymbol{Q}$) \emph{i.e.} $c^\dagger_{\boldsymbol{k}, \boldsymbol{Q}, l} \equiv c^\dagger_{\boldsymbol{k}+\boldsymbol{Q}, l}$ we obtain,
\begin{equation}
H_V = \sum_{l}\sum_{\boldsymbol{k}}\sum_{\boldsymbol{Q}, \boldsymbol{Q}'} \sum_{\substack{\eta \in \{1, -1\} \\ j\in \{1, 2, 3\}}} (\delta_{\boldsymbol{Q}, \boldsymbol{Q}' - \eta\boldsymbol{g}_j} V_{l, \eta\boldsymbol{g}_j}) c^\dagger_{\boldsymbol{k}, \boldsymbol{Q}, l} c_{\boldsymbol{k}, \boldsymbol{Q}', l}.\label{eq:moireHam}
\end{equation}
The sum over $\boldsymbol{Q},\boldsymbol{Q}' $ is over \emph{all} moiré reciprocal lattice vectors, \emph{i.e.} not restricted to the first ring about the mBZ. We obtain the full moire Hamiltonian,
\begin{equation}
\begin{aligned}
H_M = &H_0 + H_V = \\ &\sum_{\boldsymbol{k} \in \mathrm{mBZ}} \sum_{\boldsymbol{Q}, \boldsymbol{Q}'}\sum_{l, l'}h^{(m)}_{(\boldsymbol{Q},l), (\boldsymbol{Q}', l')} (\boldsymbol{k}) c^\dagger_{\boldsymbol{k}, \boldsymbol{Q}, l} c_{\boldsymbol{k}, \boldsymbol{Q}', l'},
\end{aligned}
\label{eq:moireKinetic}
\end{equation}
where,
\begin{equation}
\begin{aligned}
&h^{(m)}_{(\boldsymbol{Q}, l), (\boldsymbol{Q}', l')}(\boldsymbol{k}) \\&= \left\{[h_{\mathrm{eff}}(\boldsymbol{k}+\boldsymbol{Q})]_{l, l'} + V_0 \delta_{l, l'}\right\}\delta_{\boldsymbol{Q}, \boldsymbol{Q}'} + \sum_{\substack{\eta \in \{1, -1\} \\ j\in \{1, 2, 3\}}}(\delta_{\boldsymbol{Q}, \boldsymbol{Q}' - \eta\boldsymbol{g}_j} V_{1}).
\label{eq:ApxMoireHamiltonian}
\end{aligned}
\end{equation}
Here the first term $h_{\mathrm{eff}}(\boldsymbol{k})$ represents the \emph{low energy} R5G without the moiré coupling. The moiré system retains $C_3$ symmetry and so there are three high symmetry points $\boldsymbol{\gamma}$, $\boldsymbol{\kappa_+}$ and $\boldsymbol{\kappa_-}$. To obtain a finite matrix size, we truncate the set of $\boldsymbol{Q}$ vectors in the moiré Hamiltonian to those within a ring around the $\boldsymbol{\gamma}$ point with $|\boldsymbol{Q}| \le 5 [2\pi \cdot (a_{\mathrm{moir\acute{e}}})^{-1}]$, where $a_{\mathrm{moir\acute{e}}}$ is the moiré lattice parameter.

\section{Chern numbers of non-interacting model}
\label{Apx:PhaseDiagramofNonIntModel}
\begin{figure}
    \centering
    \includegraphics[width = 0.8\linewidth]{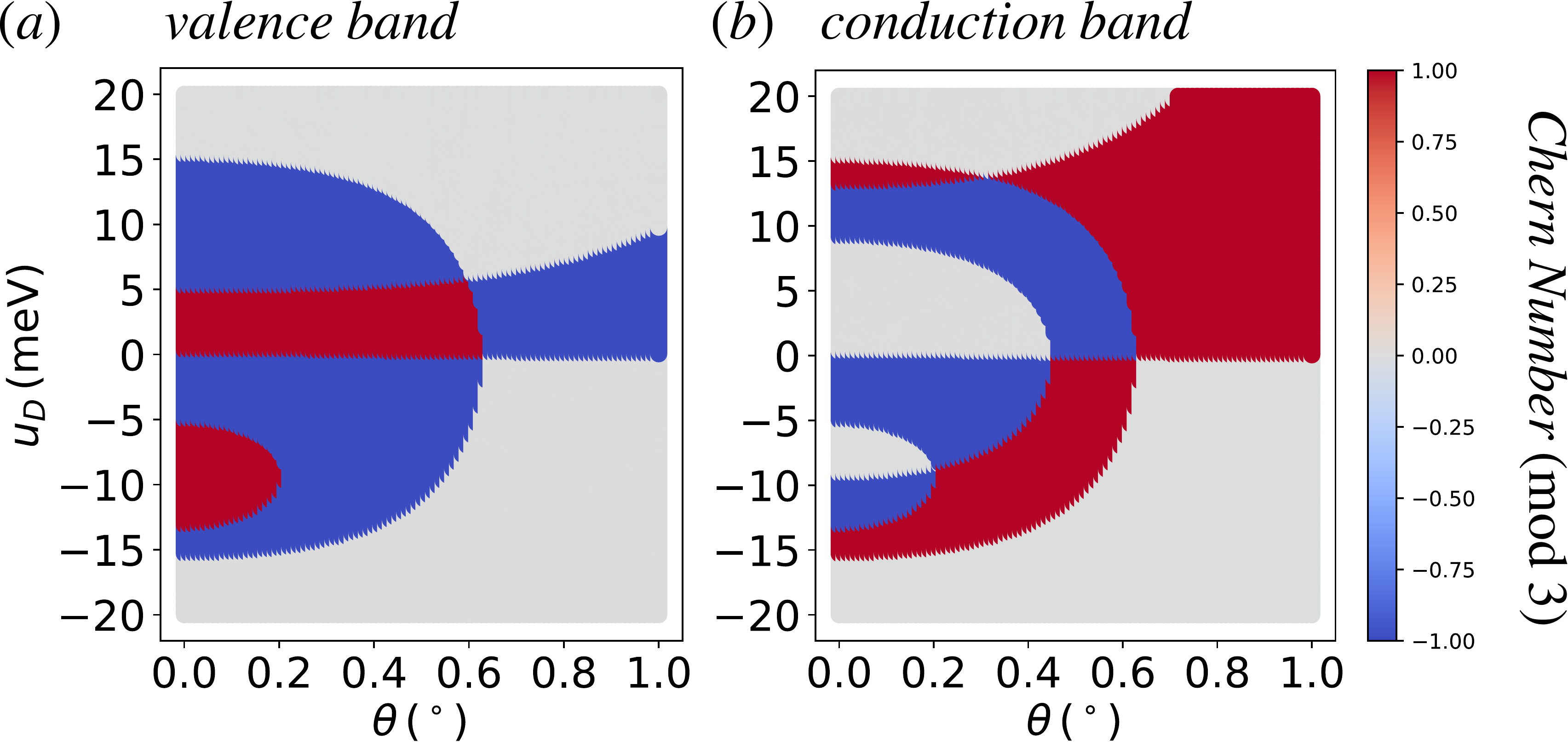}
    \caption{Chern numbers (mod 3) of the valence and conduction bands as a function of the twist angle of the hBN with respect to the R5G and the applied potential $u_D$.}
    \label{fig:ChernPhaseDiagrams}
\end{figure}

We plot the Chern numbers of the conduction and valence bands (\emph{e.g.} see Fig. 4 in main text) as a function of applied field $u_D$ and twist angle $\theta$. The resulting phase diagrams are shown in Fig.~\ref{fig:ChernPhaseDiagrams}. We see that both bands can have Chern numbers $0, 1 \:\&\:2$ (mod $3$). We choose to study excitons at the experimental angle $\theta = 0.77^\circ$ at the displacement fields $u_D = \pm20\:\mathrm{meV}$.

\begin{figure}
    \centering
    \includegraphics[width=0.75\linewidth]{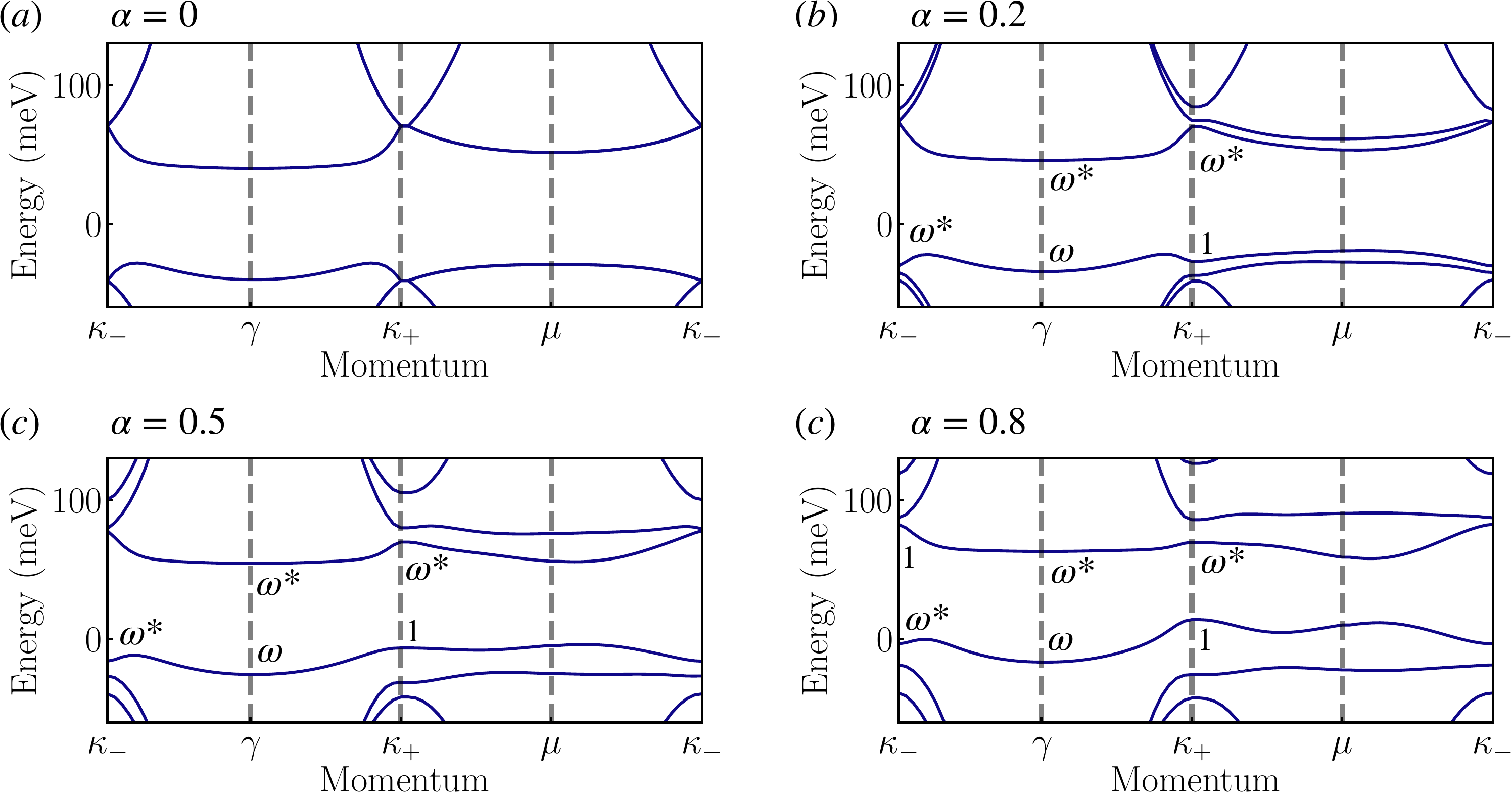}
    \caption{Tuning the moire potential to show the formation of a conduction band with Chern number 1. The $\alpha = 0$ plot shows the effects of band folding without hybridisation. When $\alpha \neq 1$ hybridisation can occur to gap out the conduction band. We plot the $C_3$ eigenvalues at the high symmetry points with $\omega = \exp\left(\mathrm{i} \frac{2\pi}{3}\right)$}
    \label{fig:gap_openingChernBand}
\end{figure}

We will now show an example of how the moire potential alters the band structure of the free-standing rhombohedral graphene model. We demonstrate this by slowing turning on the moire potential and seeing how the bands gap out to generate the Chern number 1 band at $u_D = 20 \:\mathrm{meV}$. We tune the moire potential by multiplying the moire potential strengths ($V_0, V_1$) by a parameter $\alpha$. When $\alpha =0$, the moire potential is turned off, when $\alpha = 1$ the potentials are at their full values given by Ref.~\onlinecite{MoirePotential}. At $\alpha = 0$, we plot the bands folded into the smaller mBZ (Fig.~\ref{fig:gap_openingChernBand}a). However, since there is no moire potential applied, the folded bands do not yet hybridise. Note that the states at $\kappa_+$ and $\kappa_-$ (above and below the Fermi level) are three-fold degenerate. Adding the moire potential allows these degenerate states to gap out. We see that in the conduction band, at $\kappa_+$ the state with $C_3$ eigenvalue $\omega^*$ becomes lowest in energy. The conduction band states at $\kappa_-$ remain (close to) degenerate until large $\alpha$, but by $\alpha = 0.8$, we see that the state with $C_3$ eigenvalue $1$ becomes lowest in energy. The resulting $C_3$ eigenvalues for the upper band imply that the conduction band has Chern number 1 (mod 3). We see then that the Chern number 1 band was induced by the moire potential via band folding and then hybridisation.

\section{Exciton Hamiltonian}
\label{Apx:ExcitonHamiltonian}
We calculate the exciton spectra and wave functions using the following method. We project the fully many-body Hamiltonian into the variational basis,
\begin{equation}
\ket{\boldsymbol{p}, \boldsymbol{k}} = c^\dagger_{\boldsymbol{p}+\boldsymbol{k}, c} c_{\boldsymbol{k}, v} \ket{\mathrm{GS}}.
\end{equation}
The total momentum of the above state is $\boldsymbol{p}$, the relative momentum is $\boldsymbol{k}$.

We project the full Hamiltonian onto the variational basis and calculate its matrix elements,
\begin{equation}
\bra{\boldsymbol{p}, \boldsymbol{k}} \hat H \ket{\boldsymbol{p}, \boldsymbol{k}'}.
\end{equation}
We have a contribution from the kinetic Hamiltonian [Eq.~\eqref{eq:moireKinetic}],
\begin{equation}
\bra{\boldsymbol{p}, \boldsymbol{k}} \hat H_{0} \ket{\boldsymbol{p}, \boldsymbol{k}'} = \delta_{\boldsymbol{k},\boldsymbol{k}'} (E_{\mathrm{G.S.}} - \epsilon_{\boldsymbol{k}, v} + \epsilon_{\boldsymbol{p}+\boldsymbol{k}, c}),
\end{equation}
where $\epsilon_{\boldsymbol{p}, c/v}$ are the single particle energies of the kinetic hamiltonian for the conduction/valence bands at momentum $\boldsymbol{p}$. The ground state energy is $E_{\mathrm{G.S.}} = \sum_{\boldsymbol{k}} \epsilon_{\boldsymbol{k},v}$.

We also project the interaction term into this basis. We first derive the general form of the matrix elements before showing the results for the moiré Hamiltonian. We begin with the general form of a translationally-symmetric interaction Hamiltonian,
\begin{equation}
\hat H_{int} = \sum_{\boldsymbol{R},\boldsymbol{R}',i,j} U_{\boldsymbol{R}-\boldsymbol{R}', i, j} \hat n_{\boldsymbol{R},i} \hat n_{\boldsymbol{R}',j} 
\label{eq:Hint_orig} 
\end{equation}
where $\hat n_{\boldsymbol{R},i}$ is the number operator at unit cell $R$ and $i$ labels the sublattices. Now we write this in the momentum basis. First note that
\begin{align}
c^\dagger_{\boldsymbol{k},i} &= \frac{1}{\sqrt{V}} \sum_{\boldsymbol{R}} e^{i\boldsymbol{k}\cdot\boldsymbol{R}} c^\dagger_{\boldsymbol{R},i}\\
c^\dagger_{\boldsymbol{R},i} &= \frac{1}{\sqrt{V}} \sum_{\boldsymbol{R}} e^{-i\boldsymbol{k}\cdot\boldsymbol{R}} c^\dagger_{\boldsymbol{k},i}.
\end{align}
We rewrite the number operators as,
\begin{align}
n_{\boldsymbol{R},i} &= c^\dagger_{\boldsymbol{R},i} c_{\boldsymbol{R},i}\\
 &= \frac{1}{V} \sum_{\boldsymbol{p},\boldsymbol{q}} e^{-i(\boldsymbol{p}-\boldsymbol{q})\cdot \boldsymbol{R}} c^\dagger_{\boldsymbol{p},i} c_{\boldsymbol{q},i}.
\end{align}
Equation \ref{eq:Hint_orig} becomes,
\begin{align}
\hat H_{\mathrm{int}} &= \sum_{\boldsymbol{R},\boldsymbol{R}', i, j} \sum_{\boldsymbol{p},\boldsymbol{q},\boldsymbol{p}',\boldsymbol{q}'} \frac{1}{V^2} U_{\boldsymbol{R}-\boldsymbol{R}', ij} e^{-i(\boldsymbol{p}-\boldsymbol{q})\cdot \boldsymbol{R} - i(\boldsymbol{p}'-\boldsymbol{q}')\cdot \boldsymbol{R}'} c^\dagger_{\boldsymbol{p}i} c_{\boldsymbol{q}i} c^\dagger_{\boldsymbol{p}'j} c_{\boldsymbol{q}'j}\\
 &= \sum_{\boldsymbol{R},\boldsymbol{R}', i, j} \sum_{\boldsymbol{p},\boldsymbol{q},\boldsymbol{p}',\boldsymbol{q}'} \frac{1}{V^2} U_{\boldsymbol{R}, ij} e^{-i(\boldsymbol{p}-\boldsymbol{q})R - i(\boldsymbol{p}'-\boldsymbol{q}'+\boldsymbol{p}'-\boldsymbol{q}')\cdot \boldsymbol{R}'} c^\dagger_{\boldsymbol{p}i} c_{\boldsymbol{q}i} c^\dagger_{\boldsymbol{p}'j} c_{\boldsymbol{q}'j},
\end{align}
where we set $\boldsymbol{R}-\boldsymbol{R}' \rightarrow \boldsymbol{R}$. This can be further simplified by performing the sum over $\boldsymbol{R}'$ and $\boldsymbol{q}'$ to give
\begin{equation}
\hat H_{\mathrm{int}} = \frac{1}{V} \sum_{\boldsymbol{R},i,j} \sum_{\boldsymbol{p}\boldsymbol{q}\boldsymbol{p}'} U_{\boldsymbol{R},ij} e^{-i(\boldsymbol{p}-\boldsymbol{q})\cdot \boldsymbol{R}} c^\dagger_{\boldsymbol{p},i} c_{\boldsymbol{q},i} c^\dagger_{\boldsymbol{p}',j} c_{\boldsymbol{p}-\boldsymbol{q}+\boldsymbol{p}',j}.
\end{equation}
Let $U_{\boldsymbol{p},ij} = \frac{1}{V} \sum_{\boldsymbol{R}} U_{\boldsymbol{R},ij} e^{-i\boldsymbol{p}\cdot \boldsymbol{R}}$, 
\begin{equation}
\hat H_{\mathrm{int}} = \sum_{\boldsymbol{p}\boldsymbol{q}\boldsymbol{q}'ij} U_{\boldsymbol{p},ij} c^\dagger_{\boldsymbol{p}+\boldsymbol{q},i} c_{\boldsymbol{q},i} c^\dagger_{\boldsymbol{q}',j} c_{\boldsymbol{p}+\boldsymbol{q}',j}.
\label{eq:ham_int_sublattice_basis}
\end{equation}
To calculate the matrix elements\begin{equation}
\bra{\boldsymbol{p}, \boldsymbol{k}} \hat H_{\mathrm{int}} \ket{\boldsymbol{p}, \boldsymbol{k}'}
\label{eq:int_matrix_element}
\end{equation}
we express the interaction Hamiltonian in the band basis. The Bloch function for band $\alpha$, is $u^{\boldsymbol{k},\alpha}_{i}$ meaning that,
\begin{align}
c^\dagger_{\boldsymbol{k},\alpha} &= \sum_i u^{\boldsymbol{k},\alpha}_i c^\dagger_{\boldsymbol{k},i}\\
c_{\boldsymbol{k},\alpha} &= \sum_i \bar u^{\boldsymbol{k},\alpha}_i c_{\boldsymbol{k},i}
\end{align}
where $i$ sums over the sub-lattices, and $\alpha \in \{c, v\}$. We use the notation $\bar u^{\boldsymbol{k}, \alpha}_i = (u^{\boldsymbol{k}, \alpha}_i)^*$.
Plugging this relation into equation \ref{eq:ham_int_sublattice_basis} we obtain
\begin{align}
\hat H_{\mathrm{int}} &= \sum_{\boldsymbol{p}\boldsymbol{q}\boldsymbol{q}'ij} U_{\boldsymbol{p},ij} c^\dagger_{\boldsymbol{p}+\boldsymbol{q},i} c_{\boldsymbol{q},i} c^\dagger_{\boldsymbol{q}',j} c_{\boldsymbol{p}+\boldsymbol{q}',j}\\
 &= \sum_{\boldsymbol{p}\boldsymbol{q}\boldsymbol{q}'ij\alpha\alpha'\beta\beta'} U_{\boldsymbol{p},ij} \bar{u}^{\boldsymbol{p}+\boldsymbol{q}, \alpha}_i u^{\boldsymbol{q}\alpha'}_{i} \bar{u}^{\boldsymbol{q}', \beta}_j u^{\boldsymbol{p}+\boldsymbol{q}'\beta'}_{j} c^\dagger_{\boldsymbol{p}+\boldsymbol{q},\alpha}   c_{\boldsymbol{q},\alpha'} c^\dagger_{\boldsymbol{q}',\beta} c_{\boldsymbol{p}+\boldsymbol{q}',\beta'}.
\label{eq:ham_int_band_basis}
\end{align}
Defining $U^{\boldsymbol{p}\boldsymbol{q}\boldsymbol{q}'}_{\alpha\alpha'\beta\beta'} = \sum_{ij} (U_{\boldsymbol{p},ij} \bar{u}^{\boldsymbol{p}+\boldsymbol{q}, \alpha}_i u^{\boldsymbol{q}\alpha'}_{i} \bar{u}^{\boldsymbol{q}', \beta}_j u^{\boldsymbol{p}+\boldsymbol{q}'\beta'}_{j}) $ we can write
\begin{equation}
\hat H_{\mathrm{int}} = \sum_{\boldsymbol{p}\boldsymbol{q}\boldsymbol{q}'\alpha\alpha'\beta\beta'} U^{\boldsymbol{p}\boldsymbol{q}\boldsymbol{q}'}_{\alpha\alpha'\beta\beta'} c^\dagger_{\boldsymbol{p}+\boldsymbol{q},\alpha}   c_{\boldsymbol{q},\alpha'} c^\dagger_{\boldsymbol{q}',\beta} c_{\boldsymbol{p}+\boldsymbol{q}',\beta'} .
\end{equation}
The matrix elements are,
\begin{equation}
\bra{\boldsymbol{p}, \boldsymbol{k}} \hat H_{\mathrm{int}} \ket{\boldsymbol{p}, \boldsymbol{k}'} = \sum_{\boldsymbol{l}\boldsymbol{q}\boldsymbol{q}'\alpha\alpha'\beta\beta'} U^{\boldsymbol{l}\boldsymbol{q}\boldsymbol{q}'}_{\alpha\alpha'\beta\beta'} \bra{\mathrm{GS}} c^\dagger_{\boldsymbol{k},v} c_{\boldsymbol{p}+\boldsymbol{k},c} c^\dagger_{\boldsymbol{l}+\boldsymbol{q},\alpha}   c_{\boldsymbol{q},\alpha'} c^\dagger_{\boldsymbol{q}',\beta} c_{\boldsymbol{l}+\boldsymbol{q}',\beta'} c^\dagger_{\boldsymbol{p}+\boldsymbol{k}',c} c_{\boldsymbol{k}',v} \ket{\mathrm{GS}}.
\label{eq:final_h_int}
\end{equation}
Wick's theorem can be used to simplify the expectation value,
\begin{equation}
\bra{\mathrm{GS}} c^\dagger_{\boldsymbol{k},v} c_{\boldsymbol{p}+\boldsymbol{k},c} c^\dagger_{\boldsymbol{l}+\boldsymbol{q},\alpha}   c_{\boldsymbol{q},\alpha'} c^\dagger_{\boldsymbol{q}',\beta} c_{\boldsymbol{l}+\boldsymbol{q}',\beta'} c^\dagger_{\boldsymbol{p}+\boldsymbol{k}',c} c_{\boldsymbol{k}',v} \ket{\mathrm{GS}}.
\end{equation}
For readability we simplify the notation to
\begin{align}
\mu^\dagger &= c^\dagger_{\boldsymbol{k},v},\\
\nu &= c_{\boldsymbol{p}+\boldsymbol{k},c},\\
\zeta^\dagger &= c^\dagger_{\boldsymbol{l}+\boldsymbol{q}, \alpha},\\
\sigma &= c_{\boldsymbol{q},\alpha'},\\
\chi^\dagger &= c^\dagger_{\boldsymbol{q}',\beta},\\
\gamma &= c_{\boldsymbol{l}+\boldsymbol{q}', \beta'},\\
\lambda^\dagger &= c^\dagger_{\boldsymbol{p}+\boldsymbol{k}',c},\\
\xi &= c_{\boldsymbol{k}',v}.
\end{align}
The expectation value becomes
\begin{equation}
\langle \mu^\dagger \nu \zeta^\dagger \sigma \chi^\dagger \gamma \lambda^\dagger \xi \rangle
\end{equation}
This can be expanded using Wick's theorem as
\begin{multline}
\langle \mu^\dagger \nu \zeta^\dagger \sigma \chi^\dagger \gamma \lambda^\dagger \xi \rangle = \langle \mu^\dagger \sigma \rangle (\langle \nu \zeta^\dagger \rangle \langle \chi^\dagger \xi \rangle \langle \gamma \lambda^\dagger \rangle - \langle \nu \chi^\dagger \rangle \langle \zeta^\dagger \xi \rangle \langle \gamma \lambda^\dagger \rangle - \langle \nu \lambda^\dagger \rangle (\langle \zeta^\dagger \xi \rangle \langle \chi^\dagger \gamma \rangle - \langle \zeta^\dagger \gamma \rangle \langle \chi^\dagger \xi \rangle)) \\
+ \langle \mu^\dagger \gamma \rangle (- \langle\nu \zeta^\dagger \rangle \langle \sigma \lambda^\dagger\rangle\langle \chi^\dagger \zeta \rangle + \langle \nu \chi^\dagger \rangle \langle \zeta^\dagger \xi \rangle \langle \sigma \lambda^\dagger \rangle - \langle v \lambda^\dagger \rangle (\langle \zeta^\dagger \sigma \rangle \langle \chi^\dagger \xi \rangle + \langle \zeta^\dagger \xi \rangle \langle \sigma \chi^\dagger \rangle))\\
+ \langle \mu^\dagger \xi\rangle ( \langle \nu \zeta^\dagger \rangle (\langle \sigma \chi^\dagger \rangle \langle \gamma \lambda^\dagger \rangle + \langle \sigma \lambda^\dagger \rangle \langle \chi^\dagger \gamma \rangle ) + \langle \nu \chi^\dagger \rangle ( \langle \zeta^\dagger \sigma \rangle \langle \gamma \lambda^\dagger \rangle - \langle \zeta^\dagger \gamma \rangle \langle \sigma \lambda^\dagger \rangle ) + \\ \langle \nu \lambda^\dagger \rangle ( \langle \zeta^\dagger \sigma \rangle \langle \chi^\dagger \gamma \rangle + \langle \zeta^\dagger \gamma \rangle \langle \sigma \chi^\dagger \rangle)).
\end{multline}

We simplify the expression using the values of the two particle correlators. After performing the sum in equation \ref{eq:final_h_int} we obtain,
\begin{multline}
\bra{\boldsymbol{p}, \boldsymbol{k}} \hat H_{\mathrm{int}} \ket{\boldsymbol{p}, \boldsymbol{k}'} = U^{\boldsymbol{p},\boldsymbol{k},\boldsymbol{k}'}_{cvvc} + U^{-\boldsymbol{p},\boldsymbol{p}+\boldsymbol{k}',\boldsymbol{p}+\boldsymbol{k}}_{vccv} - U^{\boldsymbol{k}'-\boldsymbol{k},\boldsymbol{k},\boldsymbol{p}+\boldsymbol{k}}_{vvcc} - U^{\boldsymbol{k}-\boldsymbol{k}',\boldsymbol{p}+\boldsymbol{k}',\boldsymbol{k}'}_{ccvv}\\ + \delta_{\boldsymbol{k},\boldsymbol{k}'} \sum_{\boldsymbol{q}} ( U^{\boldsymbol{q},\boldsymbol{k},\boldsymbol{k}}_{vvvv} - U^{\boldsymbol{0},\boldsymbol{k},\boldsymbol{q}}_{vvvv} - U^{\boldsymbol{0},\boldsymbol{q},\boldsymbol{k}}_{vvvv} -U^{\boldsymbol{q},\boldsymbol{k}-\boldsymbol{q},\boldsymbol{k}-\boldsymbol{q}}_{vccv} + U^{\boldsymbol{p}+\boldsymbol{k}-\boldsymbol{q},\boldsymbol{q},\boldsymbol{q}}_{cccc} + U^{\boldsymbol{0},\boldsymbol{p}+\boldsymbol{k},\boldsymbol{q}}_{ccvv} + U^{\boldsymbol{0},\boldsymbol{q},\boldsymbol{p}+\boldsymbol{k}}_{vvcc} - U^{\boldsymbol{q},\boldsymbol{p}+\boldsymbol{k},\boldsymbol{p}+\boldsymbol{k}}_{vccv})\\ + \delta_{\boldsymbol{k},\boldsymbol{k}'} \sum_{\boldsymbol{q},\boldsymbol{q}'} (U^{\boldsymbol{0}, \boldsymbol{q}, \boldsymbol{q}'}_{vvvv} + U^{\boldsymbol{q}, \boldsymbol{q}', \boldsymbol{q}'}_{vccv})\label{eq:InteractingHam}
\end{multline}
Finally we note that the final term is equal to the shift of the ground state energy due to the interaction \emph{i.e.}
\begin{equation}
\bra{\mathrm{GS}} H_{\mathrm{int}}\ket{\mathrm{GS}} = \sum_{\boldsymbol{q},\boldsymbol{q}'} (U^{\boldsymbol{0}, \boldsymbol{q}, \boldsymbol{q}'}_{vvvv} + U^{\boldsymbol{q}, \boldsymbol{q}', \boldsymbol{q'}}_{vccv}).
\end{equation}
Therefore to find the exciton spectrum compared to the modified ground state energy, we subtract this term from the exciton Hamiltonian. 

For a moiré Hamiltonian [\emph{e.g.} of the form Eq.~\eqref{eq:ApxMoireHamiltonian}] the $U$ matrices become,
\begin{equation} U^{\boldsymbol{p}, \boldsymbol{k}, \boldsymbol{k}'}_{\alpha, \alpha', \beta, \beta'} = \sum_{\boldsymbol{Q}} V(\boldsymbol{p}+\boldsymbol{Q}) \braket{u_{\boldsymbol{p}+\boldsymbol{k}+\boldsymbol{Q}}^\alpha|u_{\boldsymbol{p}}^{\alpha'}}\braket{u_{\boldsymbol{p}'}^{\beta}|u_{\boldsymbol{p}+\boldsymbol{k}'+\boldsymbol{Q}}^{\beta'}}.
\label{eq:BlochProducts},\end{equation}
where $\ket{u_{\boldsymbol{p}}^{\alpha}}$ is the eigenstate of the moiré Hamiltonian [Eq.~\eqref{eq:ApxMoireHamiltonian}] with momentum $\boldsymbol{p}$ and for band $\alpha \in \{c, v\}$. The interactions are modelled as a double-gated Coulomb potential as in Ref.~\onlinecite{VishwanathPaper},
\begin{equation} V(\boldsymbol{p}) = \frac{V_0 \tanh(|\boldsymbol{p}|d)}{|\boldsymbol{p}|}, \end{equation}
where $d$ is the gate distance. We set $d = 250\: \text{\r{A}}$~\cite{VishwanathPaper} for the following calculations. The interaction strength $V_0 = \frac{2\pi}{\epsilon_0 \epsilon_r}$ depends on the relative permittivity $\epsilon_r$, for which we use $\epsilon_r = 6$. Note that the expression is in Gaussian units. 

\section{Exciton Wannier states}
    \label{Apx:ExcitonWannerStates}
\begin{figure}
    \centering
    \includegraphics[width=0.5\linewidth]{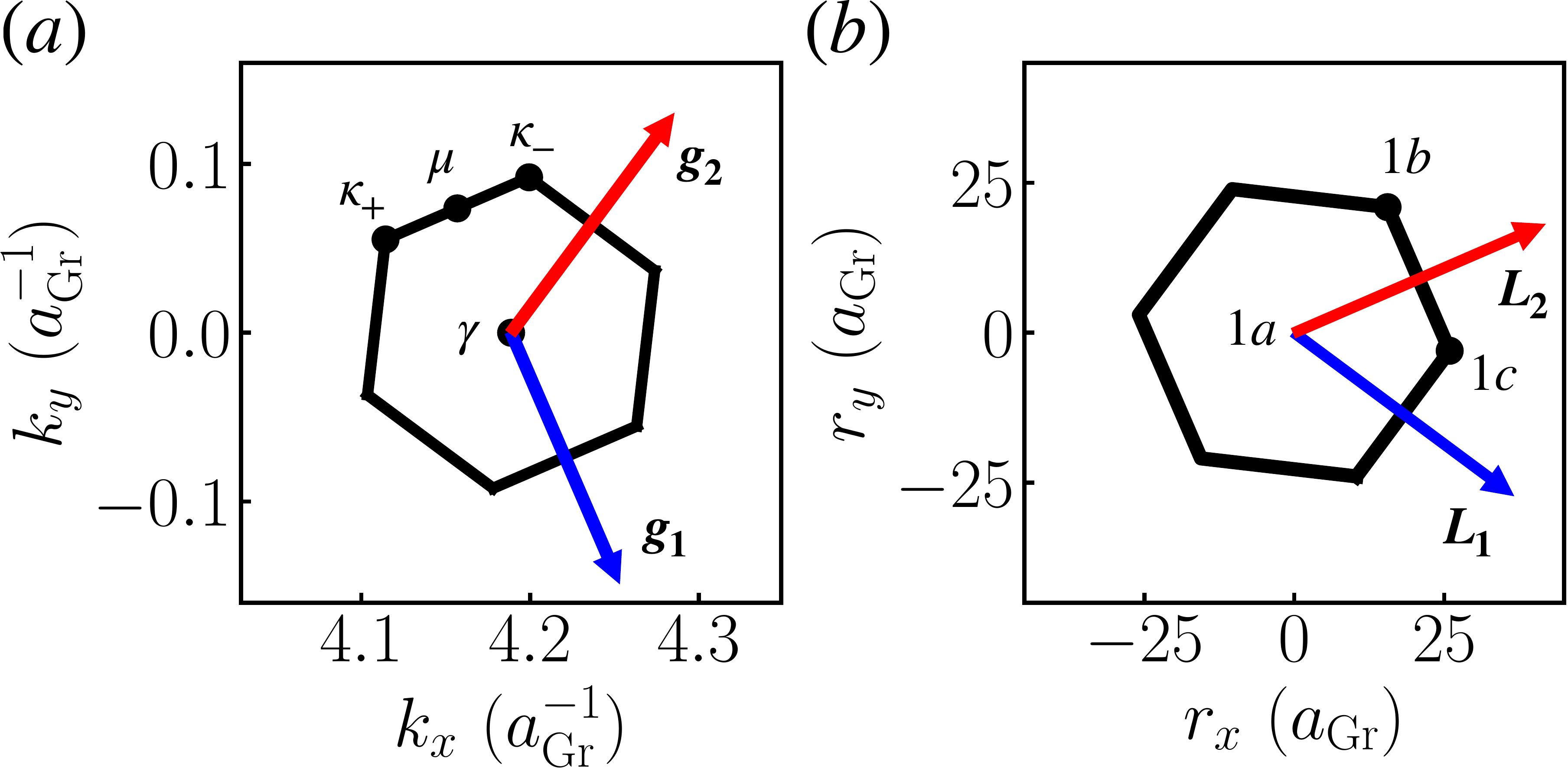}
    \caption{Panel $(a)$ is the mBZ of hBN/R5G/hBN for $\theta = 0.77^\circ$, moiré reciprocal lattice vectors $\boldsymbol{g_1}, \:\boldsymbol{g_2}$, and high symmetry points $\boldsymbol{\gamma}$, $\boldsymbol{\kappa_+}$, $\boldsymbol{\kappa_-}$ and $\boldsymbol{\mu}$. The real space unit cell [panel $(b)$] shows the lattice vectors $\boldsymbol{L_1}, \: \boldsymbol{L_2}$ and maximal Wyckoff positions $1a,\: 1b, \: 1c$.}
    \label{fig:mBZandUnitCellApe}
\end{figure}

\begin{table}[h]
    \centering
    \begin{tabular}{c|cccc}
        \hline
        $u_D$ & $\lambda^{\mathrm{exc}}_{\gamma}$ & $\lambda^{\mathrm{exc}}_{\kappa_+}$ & $\lambda^{\mathrm{exc}}_{\kappa_-}$ & $\boldsymbol{s}_{\mathrm{exc}}$\\
        \hline
        $20$ meV & $\omega$ & $1$ & $\omega^*$ & $1c$\\
        $-20$ meV & $\omega^*$ & $1$ & $\omega$ & $1b$ \\
        \hline
    \end{tabular}
    \caption{\( C_3 \) eigenvalues of the exciton wave function at high-symmetry points in the Brillouin zone for two different parameter choices. The exciton Wannier centre $\boldsymbol{s}_{\mathrm{exc}}$ is stated in terms of the Wyckoff positions (see Fig.~\ref{fig:mBZandUnitCellApe}b).}
    \label{tab:C3eigenvaluesApe}
\end{table}
We now use symmetry indicators to calculate the exciton Wannier centres. We label the exciton eigenstates in the exciton band of interest as $\ket{\psi^{\boldsymbol{p}}}$. The maximally localised exciton Wannier states then take the form,
\begin{equation}
\ket{W^{\boldsymbol{R}}} = \frac{1}{\sqrt{V}} \sum_{\boldsymbol{p}} e^{-\mathrm{i} \boldsymbol{p} \cdot \boldsymbol{R}} e^{\mathrm{i} \phi(\boldsymbol{p})} \ket{\psi^{\boldsymbol{p}}}
\end{equation}
where $e^{\mathrm{i} \phi(\boldsymbol{p})}$ is a gauge transformation that maximally localises the exciton Wannier state~\cite{davenport2025excitonberryology, jonahMLXWF}. The $C_3$ eigenvalues of the exciton bands are shown in Tab~\ref{tab:C3eigenvaluesApe}. Since the exciton band is trivial, the exciton Wannier states are exponentially localised. The $C_3$ symmetry then requires that the exciton Wannier centres sit at one of the maximal Wyckoff positions ($1a, 1b, 1c$ in Fig.~\ref{fig:mBZandUnitCellApe}b). We calculate the $C_3$ eigenvalues we expect for exciton Wannier states localised at each of these locations. 

\subsection{Wyckoff position: $1a$}
The Wyckoff position 1a lies at the centre of the moiré unit cell. We wish to determine the relationship between the exciton Wannier centre and the exciton $C_3$ eigenvalues. We first express our exciton wave function in terms of the maximally localised Wannier functions,
\begin{equation}
 \ket{\psi^{\boldsymbol{p}}}= \frac{1}{\sqrt{V}} \sum_{\boldsymbol{p}} e^{\mathrm{i} \boldsymbol{p} \cdot \boldsymbol{R}}\ket{W^{\boldsymbol{R}}}.
\end{equation}
Next we consider how the $C_3$ operator acts on this wave function. The $C_3$ operator rotates real space positions by $2\pi/3$ about the $1a$ Wyckoff position within the $\boldsymbol{R} = 0$ unit cell. Therefore, under $C_3$, the exciton Wannier state for unit cell $\boldsymbol{R}$ is mapped to unit cell $U_{C_3}\boldsymbol{R}$ (the $U_{C_3}$ operator rotates real space positions by $2\pi/3$). Applying $\hat C_3$ to the exciton wave function gives,
\begin{align}
\hat C_3 \ket{\psi^{\boldsymbol{p}}} &= \frac{1}{\sqrt{V}} \sum_{\boldsymbol{R}} e^{\mathrm{i} \boldsymbol{p} \cdot \boldsymbol{R}}\hat C_3\ket{W^{\boldsymbol{R}}}\\
 &=\lambda_{W}\frac{1}{\sqrt{V}} \sum_{\boldsymbol{R}} e^{\mathrm{i} \boldsymbol{p} \cdot \boldsymbol{R}}\ket{W^{\hat U_{C_3}\boldsymbol{R}}}\\
  &=\lambda_{W}\frac{1}{\sqrt{V}} \sum_{\boldsymbol{R}} e^{\mathrm{i} \boldsymbol{p} \cdot \hat U_{C_3}^{-1}\boldsymbol{R}}\ket{W^{\boldsymbol{R}}}\\
    &=\lambda_{W}\frac{1}{\sqrt{V}} \sum_{\boldsymbol{R}} e^{\mathrm{i} \hat U_{C_3}\boldsymbol{p} \cdot \boldsymbol{R}}\ket{W^{\boldsymbol{R}}},
\end{align}
where $\lambda_{W}$ is a phase such that $\hat C_3 \ket{W^{\boldsymbol{R}}} =\lambda_{W} \ket{W^{U_{C_3}\boldsymbol{R}}}$. Since $[\hat C_{3}]^3$ is the identity therefore $\lambda_{W}\in \{1, \omega, \omega^*\}$. At the high symmetry points $\boldsymbol{\tilde p} \in \{\boldsymbol{\gamma}, \boldsymbol{\kappa}_+, \boldsymbol{\kappa}_-\}$ (see Fig.~\ref{fig:mBZandUnitCellApe}) the exciton wave function is an eigenfunction of the $\hat C_3$ operator,
\begin{align}
\hat C_3 \ket{\psi^{\boldsymbol{\tilde p}}} &=\lambda_{W}\frac{1}{\sqrt{V}} \sum_{\boldsymbol{ R}} e^{\mathrm{i} \hat U_{C_3}\boldsymbol{\tilde p} \cdot \boldsymbol{R}}\ket{W^{\boldsymbol{R}}}\\
&=\lambda_{W}\frac{1}{\sqrt{V}} \sum_{\boldsymbol{R}} e^{\mathrm{i} \boldsymbol{p} \cdot \boldsymbol{R}}\ket{W^{\boldsymbol{R}}}\\
&=\lambda_{W}\ket{\psi^{\boldsymbol{\tilde p}}}.
\end{align}
We see therefore that, when the exciton Wannier centres are at the $1a$ Wyckoff position, the $C_3$ eigenvalues are the same at all high symmetry points (and equal to $\lambda_W$). This is not true for either of the bands in Tab~\ref{tab:C3eigenvaluesApe} and so their exciton Wannier centres are not at the $1a$ Wyckoff position. 

\subsection{Wyckoff position: $1b$}
We now find the symmetry eigenvalues of the exciton wave function if the exciton Wannier states sit at Wyckoff position $1b$ (Fig~\ref{fig:mBZandUnitCellApe}). The $\hat C_3$ operator rotates about the $1a$ Wyckoff position. Therefore, when the exciton Wannier state is centred at $1b$, the exciton Wannier centre is transformed $\boldsymbol{R}\rightarrow \hat U_{C_3}\boldsymbol{R} - \boldsymbol{L}_2$ under the action of $\hat C_3$ (see Fig.~\ref{fig:mBZandUnitCellApe}b).

We consider the action of the $C_3$ operator on the exciton wave function,
\begin{align}
\hat C_3 \ket{\psi^{\boldsymbol{p}}} &= \frac{1}{\sqrt{V}} \sum_{\boldsymbol{R}} e^{\mathrm{i} \boldsymbol{p} \cdot \boldsymbol{R}}\hat C_3\ket{W^{\boldsymbol{R}}}\\
 &=\lambda_{W}\frac{1}{\sqrt{V}} \sum_{\boldsymbol{R}} e^{\mathrm{i} \boldsymbol{p} \cdot \boldsymbol{R}}\ket{W^{\hat U_{C_3}\boldsymbol{R} - \boldsymbol{L}_2}}\\
  &=\lambda_{W}\frac{1}{\sqrt{V}} \sum_{\boldsymbol{R}} e^{\mathrm{i} \boldsymbol{p} \cdot \hat U_{C_3}^{-1}(\boldsymbol{R}+\boldsymbol{L}_2)}\ket{W^{\boldsymbol{R}}}\\
  &=\lambda_{W}\frac{1}{\sqrt{V}} \sum_{\boldsymbol{R}} e^{\mathrm{i}  \hat U_{C_3} \boldsymbol{p} \cdot(\boldsymbol{R}+\boldsymbol{L}_2)}\ket{W^{\boldsymbol{R}}}.
\end{align}
Hence at the high symmetry points $\boldsymbol{\tilde p}$,
\begin{align}
\hat{C}_3 \ket{\psi^{\boldsymbol{\tilde{p}}}} 
&= \lambda_{W} \frac{1}{\sqrt{V}} \sum_{\boldsymbol{R}} 
e^{\mathrm{i} \boldsymbol{\tilde{p}} \cdot (\boldsymbol{R} + \boldsymbol{L}_2)} 
\ket{W^{\boldsymbol{R}}}\\
&= \left(\lambda_{W} e^{\mathrm{i}\boldsymbol{\tilde p}\cdot\boldsymbol{L}_2}\right)\ket{\psi^{\boldsymbol{\tilde p}}}.
\end{align}
The $C_3$ eigenvalues at the high symmetry points are therefore,
\begin{equation}
\lambda_{C_3}(\boldsymbol{\tilde p}) = \lambda_{W} e^{\mathrm{i}\boldsymbol{\tilde p}\cdot\boldsymbol{L}_2}.
\end{equation}
Using $\omega = e^{-\mathrm{i} \frac{2\pi}{3}}$ we can write the $C_3$ eigenvalues of the exciton wave function at the high symmetry points as $\lambda_{C_3}(\boldsymbol{\gamma}) = \lambda_{W}$, $\lambda_{C_3}(\boldsymbol{\kappa}_+) = \lambda_{W} \omega$ and $\lambda_{C_3}(\boldsymbol{\kappa}_-) = \lambda_{W} \omega^*$. For $\lambda_W = \omega^*$ we see that this gives identical symmetry indicators to those shown for the excitons at $u_D = -20 \:\mathrm{meV}$ (see Tab.~\ref{tab:C3eigenvaluesApe}). Therefore the exciton Wannier centres at $u_D = -20 \:\mathrm{meV}$ are at the $1b$ Wyckoff position.
\subsection{Wyckoff position: $1c$}
The same method can be used to calculate the symmetry eigenvalues for exciton Wannier states that sit at the $1c$ Wyckoff position. Here we find that under $C_3$ the exciton Wannier state at unit cell $\boldsymbol{R}$ transforms to $\hat U_{C_3} \boldsymbol{R} - \boldsymbol{L}_1$. The symmetry eigenvalues therefore are, 
\begin{equation}
\lambda_{C_3}(\boldsymbol{\tilde p}) = \lambda_{W} e^{\mathrm{i}\boldsymbol{\tilde p}\cdot\boldsymbol{L}_1}.
\end{equation}
We find that $\lambda_{C_3}(\boldsymbol{\gamma}) = \lambda_{W}$, $\lambda_{C_3}(\boldsymbol{\kappa}_+) = \lambda_{W} \omega^*$ and $\lambda_{C_3}(\boldsymbol{\kappa}_-) = \lambda_{W} \omega$. For $\lambda_{W} = \omega$ this gives the symmetry indicators we found for the exciton band at displacement field $u_D = 20\: \mathrm{meV}$ (see Tab.~\ref{tab:C3eigenvaluesApe}). Therefore the exciton Wannier states at this displacement field are centred at the Wyckoff position $1c$.

\section{Exciton Berry curvature}
\label{Apx:ExcitonBerryCurvature}
In Ref.~\onlinecite{davenport2025excitonberryology} we derived two exciton Wilson loops which relate to the projection of the electron and hole position operator into the exciton band. For 1D systems we obtain the two exciton Wilson loops for the conduction ($c$) and valence bands ($v$) respectively,
\begin{align}
W_{\mathrm{exc}, c} &= \prod_p \big[\sum_k  \bar \phi^{p+\Delta}_{k- p} \phi^{p}_{k- p} \braket{u_{k+\Delta, c}| u_{k, c}}]\label{eq:WilsonloopElec}\\
W_{\mathrm{exc}, v} &= \prod_p \big[\sum_k \bar \phi^{p+\Delta}_{k} \phi^{p}_{k+\Delta} \braket{u_{k+\Delta, v}| u_{k, v}}].
\label{eq:WilsonloopHole}
\end{align}
with $\Delta = 2\pi/L$ for system size $L$. We note that these can both be written as a product of projectors,
\begin{equation}
W_{\mathrm{exc}, c/v} = \bra{\mathcal{U}^{\mathrm{exc}, c/v}_{p_1}} \lim_{R\rightarrow\infty} \prod^2_{i = R} \ket{\mathcal{U}^{\mathrm{exc}, c/v}_{p_i}} \bra{\mathcal{U}^{\mathrm{exc}, c/v}_{p_i}}\ket{\mathcal{U}^{\mathrm{exc}, c/v}_{p_1}}
\end{equation}
using,
\begin{align}
\ket{\mathcal{U}^{\mathrm{exc}, c}_{p}} &= \sum_{k} \phi^p_k \ket{k} \otimes \ket{u_{p+k, c}} \otimes \ket{\bar u_{k, v}}\label{eq:Wavefunctions4WilsonLoops1}
\\
\ket{\mathcal{U}^{\mathrm{exc}, v}_{p}} &= \sum_{k} \phi^p_k \ket{p+k} \otimes \ket{u_{p+k, c}} \otimes \ket{\bar u_{k, v}}.
\label{eq:Wavefunctions4WilsonLoops2}
\end{align}
See Ref.~\onlinecite{davenport2025excitonberryology} for details of the derivations. 

Following the method introduced by Ref.~\onlinecite{discreteBZ_ChernNumbers} we can use these Wilson loop expressions to construct expressions for the flux through loops in the BZ. We consider the exciton wave functions which we have calculated on a grid with spacing $\boldsymbol{d}_{\mu}$ in direction $\mu$. We can define the link variables,
\begin{equation}
U_{\mu, c/v}(\boldsymbol{k}) = \frac{\braket{\mathcal{U}^{\mathrm{exc}, c/v}(\boldsymbol{k})|\mathcal{U}^{\mathrm{exc}, c/v}(\boldsymbol{k}+\boldsymbol{d}_\mu)}}{|\braket{\mathcal{U}^{\mathrm{exc}, c/v}(\boldsymbol{k})|\mathcal{U}^{\mathrm{exc}, c/v}(\boldsymbol{k}+\boldsymbol{d}_\mu)}|}.
\end{equation}
We then calculate the lattice field strength,
\begin{equation}
F_{c/v}(\boldsymbol{k}) = - \mathrm{i} \ln \left[U_{1,c/v}(\boldsymbol{k}+\boldsymbol{d}_{1})U_{2, c/v}(\boldsymbol{k}+\boldsymbol{d}_{2}) U_{1,c/v}(\boldsymbol{k})^{-1} U_{2,c/v}(\boldsymbol{k})^{-1}  \right].
\end{equation}
For small $|\boldsymbol{d}_{\mu}|$, $F_{c/v}(\boldsymbol{k})$ is proportional to the Berry curvature~\cite{discreteBZ_ChernNumbers} and the Chern number of the band can be calculated as,
\begin{equation}
C = \frac{1}{2\pi}\sum_{\boldsymbol{k}}F_{c/v}(\boldsymbol{k}).
\end{equation}
This gives us a method to calculate the exciton Berry curvature without having to first find a smooth gauge. We have two expressions for the Berry curvature but for the system studied in this paper they both give similar Berry curvature distributions. In addition, as expected they both integrate to the same value and so give the same Chern number~\cite{davenport2025excitonberryology}. In the main text we therefore only plot one of the possible Berry curvature plots - that calculated using $\ket{\mathcal{U}^{\mathrm{exc}, c}_p}$.

\end{document}